\documentclass[journal]{IEEEtran}
\usepackage{graphicx}
\usepackage{amsmath}

\usepackage{amssymb}
\usepackage[ruled]{algorithm2e}

\usepackage{dblfloatfix}

\usepackage{url}

\ifCLASSINFOpdf
\else
\fi
\begin{document}
%
\title{Automatic Camera Trajectory Control with Enhanced Immersion for Virtual Cinematography}
%
%
%

\author{Xinyi~Wu,
        Haohong~Wang,~\IEEEmembership{Senior Member,~IEEE},
        and~Aggelos~K. Katsaggelos,~\IEEEmembership{Life Fellow,~IEEE}
\thanks{This work was supported by TCL Research America through Interactive Hyperstory project.}
\thanks{X. Wu and A. K. Katsaggelos are with the Department of Electrical and Computer Engineering, Northwestern University, Evanston, IL 60208 USA (e-mail:xinyiwu2019.1@u.northwestern.edu;a-katsaggelos@northwestern.edu).}
\thanks{H. Wang is with TCL Research America, San Jose, CA 95110 USA (haohong.wang@tcl.com).}
\thanks{Manuscript received April 19, 2005; revised August 26, 2015.}}

%
%

\markboth{IEEE TRANSACTIONS ON MULTIMEDIA,~Vol.~14, No.~8, August~2015}%
{Wu \MakeLowercase{\textit{et al.}}: Automatic Camera Movement Generation with Enhanced Immersion for Virtual Cinematography  }
%

\IEEEpubid{0000--0000/00\$00.00~\copyright~2021 IEEE}

\maketitle



\begin{abstract}
User-generated cinematic creations are gaining popularity as our daily entertainment, yet it is a challenge to master cinematography for producing immersive contents. Many existing automatic methods focus on roughly controlling predefined shot types or movement patterns, which struggle to engage viewers with the actor’s circumstances. Real-world cinematographic rules show that directors can create immersion by comprehensively synchronizing the camera with the actor. Inspired by this strategy, we propose a deep camera control framework that enables actor-camera synchronization in three aspects, considering frame aesthetics, spatial action, and emotional status in the 3D virtual stage. Following rule-of-thirds, our framework first modifies the initial camera placement to position the actor aesthetically. This adjustment is facilitated by a self-supervised adjustor that analyzes frame composition via camera projection. We then design a GAN model that can adversarially synthesize fine-grained camera movement based on the actor’s action and psychological state, using an encoder-decoder generator to map kinematics and emotional variables into camera trajectories. Moreover, we incorporate a regularizer to align the generated stylistic variances with specific emotional categories and intensities. The experimental results show that our proposed method yields immersive cinematic videos of high quality, both quantitatively and qualitatively. Live examples can be found in the supplementary video.

\end{abstract}


%

\IEEEpeerreviewmaketitle

\section{Introduction}

\begin{figure*}
  \centering
  \includegraphics[width=1.0\textwidth]{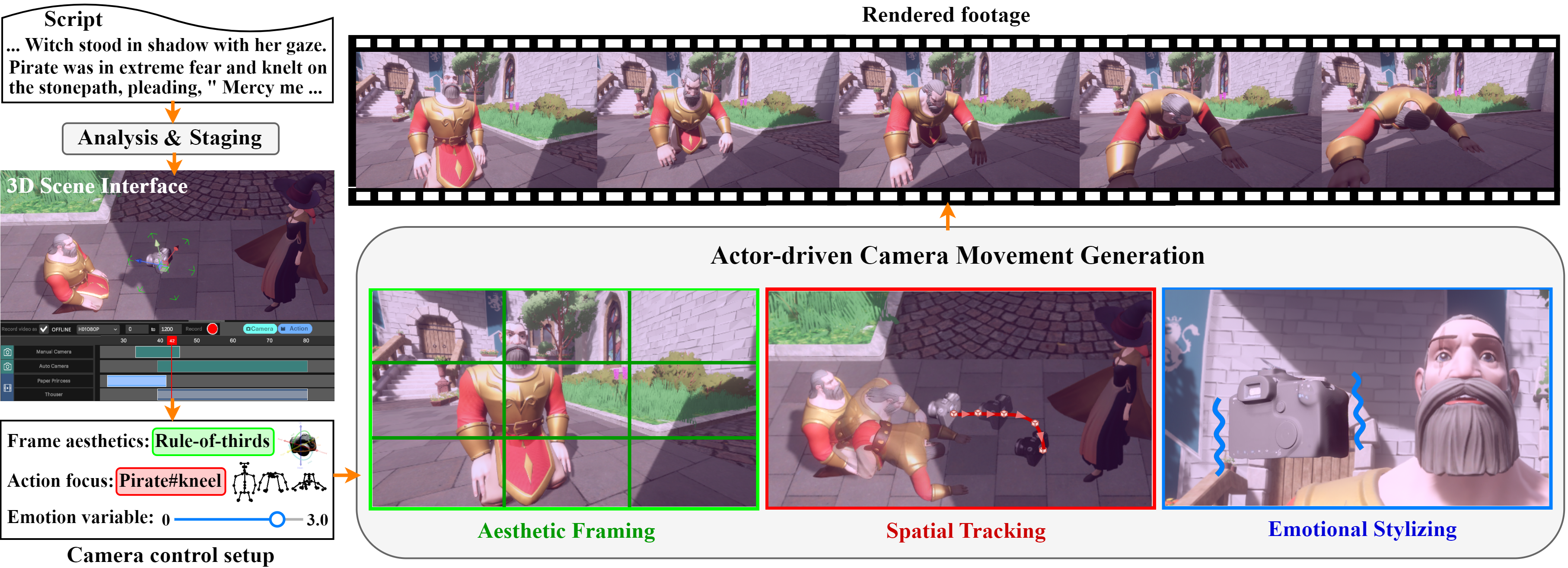}
  \caption{We propose a virtual camera controller that automates camera movements to produce footage with improved immersive experiences. This is achieved by conducting actor-camera synchronization in three key aspects: maintaining aesthetic rule-of-thirds composition, tracking the spatial action of the focused character, and stylizing the camera trajectory based on a specific emotion variable.}
\end{figure*}

\IEEEPARstart{U}{ser}-generated content (UGC) has been increasingly popular on social media platforms, with creators often adopting immersive techniques from film directors to enhance the quality and appeal of their cinematic outputs. Despite possessing interesting scripts, creators may struggle to produce engaging works due to limited cinematographic knowledge and skills, thereby also causing difficulties in creating satisfactory immersive experiences for viewers.

Cinematography is essential in developing high-quality cinematic creations. Given a certain scenario, directors follow cinematographic rules as guidance to optimize camera manipulation, which is a key component for effectively immersing the audience into the story. Such rules are defined according to combining objective aesthetic principles with subjective viewer feedback \cite{brown2016cinematography}. Although comprehensive guidelines are available for controlling camera behavior, conducting cinematography manually still remains a challenging and labor-intensive process to obtain satisfactory outputs.

Auto-cinematography has emerged as a solution to reduce labor costs and professional barriers during production. Partial answers efficiently generate film sequences via editing existing cinematic clips from multiple sources \cite{truong2016quickcut, wang2019write}. Because of lacking camera controllability, these editing-based methods create immersion only at a basic level, ensuring either spatio-temporal \cite{arev2014automatic} or semantic consistency \cite{liang2012script} for the story plot. Another feasible way is to automate camera manipulation by mimicking human directorial techniques \cite{louarn2018automated, karakostas2020shot}. Traditional camera planning methods encode specific behavior rules for varied scenarios \cite{galvane2015automatic,subramonyam2018taketoons}, yet the maintenance and extension of such encoded control patterns can result in difficulty. 

As deep neural networks dominate generative tasks, they are promising to enable flexible camera control with enhanced viewer immersion. By designing reward functions based on scene analysis, camera placements that maximize frame aesthetics can be decided via reinforcement learning \cite{gschwindt2019can,yu2022enabling}. Nevertheless, these methods overlook the crucial role of actor behavior in shaping the overall immersive experiences \cite{brown2016cinematography}. \cite{huang2019learning,huang2021one} address this via predicting camera trajectories learned from human motion videos, utilizing an end-to-end deep model. While facilitating continuous camera tracking, such long-distance drone-based approaches struggle with refined controls over detailed actor dynamics both physically and psychologically. To achieve fine-grained immersive shooting, \cite{jiang2020example,jiang2021camera} extract high-level cinematic features from footage of masterpieces and transfer them into retargetable camera rails. However, their dependency on references constrains user customization and yields implicit immersion representations. Despite the popularity in AI-based cinematography, a method that can explicitly and comprehensively immerse the audience is needed with controllable cameras \cite{wang2024enabling}.

\IEEEpubidadjcol

In real-world cinematography, directors have observed that immersion can be significantly enhanced by simulating the viewpoint of an invisible character within the scene \cite{camproxy}. Empirical studies confirm this effect is often achieved through shooting techniques—aligning camera movement with the focused actor \cite{hijackeye}—across multiple perceptual-sensitive dimensions, including aesthetics \cite{macklin2019going}, action \cite{brown2016cinematography} and emotion \cite{burelli2016game}. Therefore, to thoroughly create immersive feelings for viewers, these factors should be collectively considered.

In this paper, we propose a deep camera control framework that can enable cinematographic immersion comprehensively by satisfying actor-camera synchronization for three aspects in the 3D virtual environment. Specifically, we generate actor-driven camera trajectories to ensure aesthetic frame composition, refined spatial movement tracking, and stylistic variances in alignment with emotion variables. Our framework first leverages a self-supervised adjustment network to modify the initial camera placement following rule-of-thirds \cite{krages2012photography} for optimizing composition. Unlike previous methods that focus on scene images \cite{li2018a2,hong2021composing}, our adjustor analyzes camera projection to efficiently position the actor in a way that enhances compositional aesthetics. Subsequently, we design a GAN model to synthesize camera movements adversarially mimicking ground-truth samples from human artists. Our generator employs an encoder-decoder architecture, with the encoder capturing the key kinematic features and their saliencies from disarticulated actor poses in the input. The decoder then transforms these features into camera trajectories, conditioned on an emotion variable that indicates the emotional category and intensity of the actor. Moreover, to better align the generated stylistic variances with the given emotion variable, we incorporate a regularizer to constrain the overall shape of our synthesized camera trajectory consistent with that of manual ground truth. Both quantitative and qualitative evaluations demonstrate that our camera control framework can perform immersive shooting resembling those of professional artists and produce high-quality cinematic videos with improved viewer immersion.

The contributions of our work are summarized as follows:

\begin{itemize}
  \item We propose a deep camera control framework that learns actor-camera synchronization based on specific cinematographic techniques across three aspects: frame aesthetics, spatial action, and emotional status, to significantly enhance immersion in user-generated video creations.

\vspace{1mm}
  \item We design a self-supervised adjustor to efficiently adjust aesthetic compositions via camera projection analysis. We build a GAN model for fine-grained camera trajectory generation using actor-to-camera behavior transformation. A shape-based regularizer is further integrated to control stylistic variances in our generated trajectories.

\vspace{1mm}
  \item We conduct both quantitative and qualitative evaluations to show our superior capability of creating comprehensive immersive experiences that deeply engage the audience into physical and psychological behaviors of the actor within aesthetically appealing frames.

\end{itemize}

\vspace{2mm}

\section{Related work}

In this section, we review the progress made in auto-cinematography and explore advancements that analyze cinematographic-level factors for enhancing the immersive experience of viewers.

\subsection{Enabling automatic cinematography}

The manual production of high-quality 2D cinematic creation is costly, requiring much labor and professional knowledge. To address this, methods such as multi-source clip editing, camera behavior planning, and text-to-video generation have been proposed to automatically obtain cinematic videos. Despite significant improvements in the capability of recent LLM-based text-to-video solutions \cite{huang2024free}, they still lack stability and detailed cinematographic controls during the generation. Thus, here we primarily discuss the former two auto-cinematography methods.

\vspace{1mm}
\textbf{Multi-source clip editing:} Editing serves as a feasible approach for producing a desired cinematic video by composing various short clips. To ensure the generation of required content, \cite{liang2012script} annotated video clips with key content details, conducting retrieval through script keyword matching. \cite{arev2014automatic} took one more step to improve spatio-temporal consistency during the retrieval using graph optimization. Moreover, \cite{wang2020attention} introduced attention maps of video frames and tracked human gaze behavior to guarantee the consistent appearance of salient objects across frames. For achieving semantic consistency,  \cite{truong2016quickcut} compared script and video annotations via TF-IDF \cite{manning2009introduction} scores, whereas \cite{wang2019write} further performed text alignment by regularizing probability distribution between high-level semantic features. Instead of aligning text, \cite{yang2018text2video} focused on maintaining frame-level semantic coherence within and between clips based on constraining extracted image-based embeddings. However, the effectiveness of these editing-based methods is heavily dependent on the diversity and quality of the available clip database, potentially limiting flexibility and posing difficulties for general user-generated application scenarios.

\vspace{1mm}
\textbf{Camera behavior planning:} Another trend of auto-cinematography imitates the workflow of film directors. They first design camera control patterns and then apply these behaviors in drones or virtual cameras to create final cinematic videos. In the early stage, research in this domain performed pre-defined camera movements like pan, tilt, and zoom automatically using analytical ways based on the given script annotations \cite{hayashi2014t2v, subramonyam2018taketoons}. This ensures that the focused character can always stay within the frame. Take a further step, optimization methods have emerged to manage more complex camera behaviors, adjusting either camera intrinsics, such as focal length \cite{karakostas2020shot}, or extrinsics, from single placements \cite{louarn2018automated} to dynamic rails \cite{galvane2015automatic}. Additionally, aesthetic principles regarding frame composition \cite{pueyo2024cinempc}, actor viewpoint \cite{bonatti2020autonomous}, and action continuity \cite{yu2023novel} are incorporated as constraints during the optimization process, enhancing the overall viewing experience.

Recently, neural networks have been effective in generation tasks. \cite{gschwindt2019can} utilized Reinforcement Learning (RL) to obtain a deep camera movement agent supervised by real-time human preference scores. \cite{yu2022enabling} further designed reward functions that ensure aesthetics and fidelity, avoiding occlusion and poor shot angles following specific director rules. To offer high-quality tracking shots, \cite{ren2023automatic} introduced a visual detection network for precise camera movement guidance, while \cite{xie2023camera} leveraged transformers to track based on forming the optimal placement and orientation of the actor. Moreover, \cite{yu2023automated} focused on improving camera control in scenes with interactive actions of the characters using a GAN model.

Given that cinematic videos are widely accessible online, \cite{huang2019learning} proposed a novel approach to replicate shooting patterns in reference videos, transferring human kinematics and optical flows into reusable camera movements. This method was extended by \cite{huang2021one} and \cite{dang2022path}, who added a filming style extractor, using low-dimensional vector and RL-based path analysis, respectively, for stylistic alignment between the reference and generated footage. Yet,  \cite{huang2019learning, huang2021one,dang2022path} handle primarily long-distance aerial shots and lack detailed camera controls to meet artist-level cinematic production. To alleviate this problem, \cite{jiang2020example} created a fine-grained cinematic feature space that learns shooting patterns from film masterpieces by analyzing the inter-relationships of character poses in the frame. Building on this, \cite{jiang2021camera} introduced a keyframe control strategy to enforce stricter cinematic constraints. Similarly, \cite{wang2023jaws} refined key cinematic features in Neural Radiance Fields (NeRF) with heatmap guidance, which contributes to improved views for the output cinematic videos. Despite these advances in auto-cinematography, few camera-based methods address the sense of immersion, a critical factor of cinematic video quality that significantly influences viewer experiences.

\subsection{Investigating cinematographic immersion}

The popularity of short videos on social media drives creators to produce high-quality content that not only attracts the audience but also deeply immerses them in the story. Underscoring the  importance of visual perception in the viewer experience, research reveals that the use of cinematographic techniques such as view range manipulation \cite{jung2023immersive,horiuchi2023augmented}, staging \cite{louarn2018automated,traparic2023towards} and lighting \cite{wei2023feeling} can effectively convey the sense of immersion. Among these, camera control stands out as one of the most crucial and commonly used approaches for achieving cinematographic immersion.

In practice, directors discovered that camera behavior can benefit viewer engagement and foster resonance between actors and the audience \cite{camproxy}, particularly through precisely synchronizing the camera with the actor \cite{hijackeye}. Due to multiple factors affecting human vision, such camera-actor synchronization in manual cinematography usually involves spatial \cite{brown2016cinematography}, emotional \cite{emocam}, and aesthetic aspects \cite{macklin2019going} for creating immersive experiences thoroughly. 

Imitating human directors, auto-cinematographic methods utilized shooting techniques such as Orbiting \cite{jiang2020example,jiang2021camera} and Tracking \cite{huang2019learning,ren2023automatic} to synchronize with the actor’s physical movement. By doing so, the camera behaves in a way that the viewer immersively observes the scene, ensuring consistent presence of the actor regardless of the action. In terms of emotional resonance, Zooming \cite{karakostas2020shot} and Shaking \cite{sayed2022lookout} can handle the synchronization between the camera and the mental state of the actor, which adjust focal length and stability to reflect specific emotional categories and intensities, respectively. In addition, \cite{bonatti2021batteries} learned a semantic space for emotion representation using a large crowd-sourced footage dataset. Such a semantic space is subsequently combined during camera behavior generation to guide the synthesis of stylistic variance aligning with the input emotion type. For aesthetic-level actor-camera synchronization, the camera should follow aesthetic principles like center-framing \cite{subramonyam2018taketoons,karakostas2020shot}, the 180-degree rule \cite{yu2023novel, yu2023automated}, or the rule-of-thirds \cite{galvane2015automatic, ren2023automatic}. This allows to aesthetically immerse the audience by shooting the actor in well-designed frame compositions. There are also methods controlling frame aesthetics based on subjective user preferences \cite{gschwindt2019can,pueyo2024cinempc} or replicating styles from film masterpieces \cite{jiang2020example,jiang2021camera,wang2023jaws}.

Though various automatic camera control methods can partially address the synchronization with the actor, thereby implicitly improving the sense of immersion, there is still the need for a unified approach. This approach would systematically combine frame aesthetics, spatial action, and emotional state to explicitly solve actor-camera synchronization and produce cinematic creations that offer comprehensive immersive experiences.

\section{Method}

\begin{figure*}[htbp]
\centerline{\includegraphics[width=.98\linewidth]{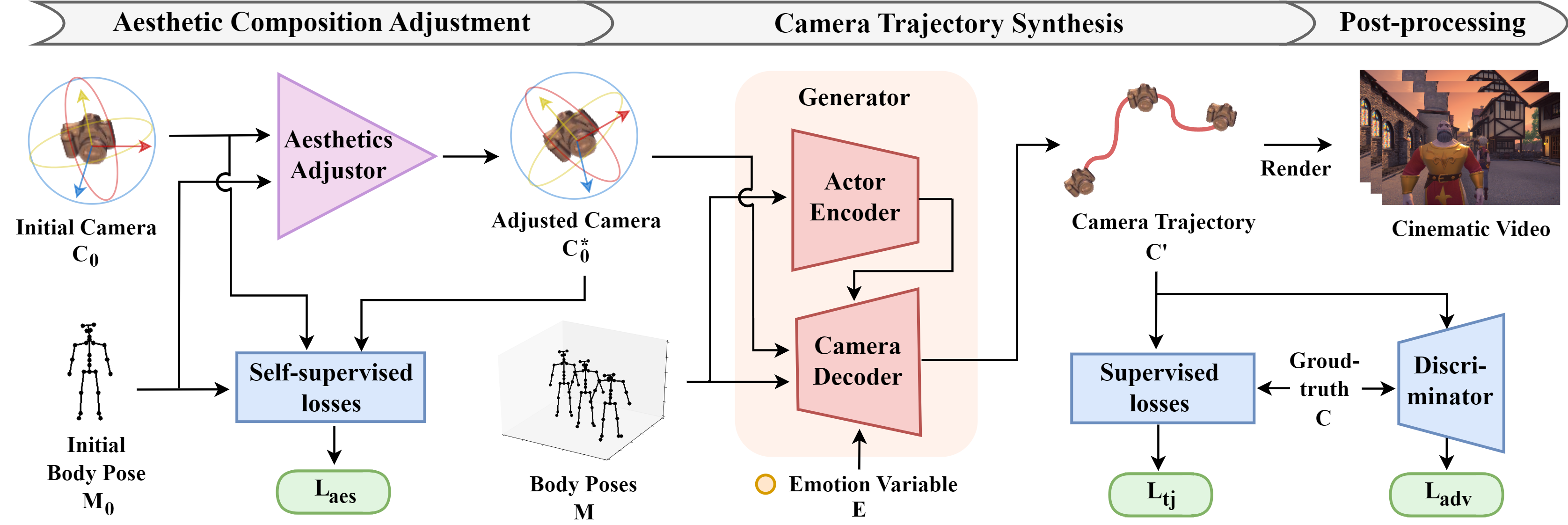}}
\caption{The overview of our proposed camera control framework, which takes user-specific data from a virtual environment to generate camera movements. Through flexible two-stage processing, it ensures actor-camera synchronization across multiple aspects for producing customized immersive cinematic videos.
}

\label{overview}
\end{figure*}

To facilitate the production of high-quality user-generated cinematic videos, we propose a novel auto-cinematography method that can handle fine-grained camera control to significantly enhance immersion of the generated works. The detailed methodology is presented in the following section.

\subsection{Immersive camera control framework}

Camera movement, especially when precisely synchronized with the actor, has been recognized by directors as one of the essentials for creating immersive feelings, thus engaging the audience in the story \cite{hijackeye}. Building on such a empirical rule, we propose a two-stage camera control framework to address actor-camera synchronization in terms of aesthetic, action, and emotional levels. This design aligns with user logic, supporting flexible and repeated applications based on individual needs, allowing the user to thoroughly enhance immersion for the output cinematic creation. Please refer to supplementary materials for more information regarding the foundational cinematographic knowledge.

In our virtual stage, users can customize their characters as well as camera placements, where all objects in the 3D environment share $Q=6$ degrees of freedom: three for the position (x, y, z) and three for the rotation (yaw, pitch, roll) axes. These specified setups are crucial in regard to automating a comprehensive synchronization between the actor and the camera. We capture the movement of the focused actor by obtaining $T$ frames of poses $M \in \mathbb{R}^{T \times JQ}$, with $J$ denoting the number of joints per pose, to enable refined spatial action tracking. To represent the mental state of the actor, a non-negative emotion variable $E$ is utilized to compactly depict both the category and its intensity for facilitating emotional styling. Meanwhile, an initial camera placement $C_0 \in \mathbb{R}^{Q}$ is required from the user so that we are able to improve frame aesthetics and initialize the generation of camera trajectories. Therefore, given such user-specific data input, our camera control framework $\Upsilon$ can be formulated as:

\begin{equation}\label{OvA}
C'=\Upsilon(M,E,C_0),
\end{equation}
where $C' \in \mathbb{R}^{T \times Q}$ represents the synthesized camera movement sequence for rendering the cinematic video.

Fig. \ref{overview} demonstrates the workflow of our camera control framework $\Upsilon$, which consists of two key modules. Realizing that users may not possess professional insights of aesthetics, we first design a deep adjustment network $\psi$, acting as an aesthetics adjustor for constructing aesthetic frame composition. Utilizing the initial camera placement $C_0$ and actor pose $M_0 \in \mathbb{R}^{JQ}$ provided by the user, the adjustor $\psi$ modifies the camera placement to $C_0^{*}$ by analyzing the actor location within the frame under camera projection. This adjustment is guided by a self-supervised hybrid loss $\mathcal{L}_{aes}$, which offers aesthetic constraints based on the rule-of-thirds principle and simultaneously minimizes potential visual changes. Note that the replacement by any other aesthetic principles does not affect the current procedure. 

Starting with a more aesthetic initialization, we employ a GAN-based deep model to generate camera movements that can synchronize with the actor’s spatial action and emotional state. Our generator $G$ adopts an encoder-decoder architecture. The encoder extracts local kinematics from the actor pose sequence $M$ and captures the hidden director tracking strategies using saliency maps. These obtained features are then transformed into a camera-space trajectory $C'$ in the decoder, conditioned on the initial camera $C_0^{*}$ and emotion variable $E$. Notably, $E$ here controls the amplitude of camera movement to match different emotional states with specific styles. The overall generation is regularized using a hybrid trajectory loss $\mathcal{L}_{tj}$, which includes point-level, shape-level, and feature-level constraints, along with an adversarial loss $\mathcal{L}_{adv}$ derived from a discriminator, compared to ground-truth human artist samples.

In this way, Equation (\ref{OvA}) can thus be reformed as: 

\begin{equation}\label{nutshell}
C'=G(M,E,\psi(M_{0},C_{0}))=G(M,E,C_0^{*}).
\end{equation}

We dive into more processing details in the next two subsections.

\subsection{Aesthetic composition adjustment}

This module modifies the user-chosen initial camera placement in order to shoot the actor in a way that achieves composition-based frame aesthetics for improving immersion. Instead of relying on image analysis \cite{li2018a2,hong2021composing}, which causes a high rendering cost, our approach conducts efficient adjustments by leveraging the 3D-to-2D camera projection following the widely-used rule-of-thirds \cite{krages2012photography} aesthetic principle.

\subsubsection{Rule-of-thirds with camera projection} The rule-of-thirds principle provides camera control guidance over various scenarios to ensure the actor is optimally positioned within the frame, forming an aesthetic frame composition. By projecting the actor's 3D pose onto the 2D frame from a given camera placement, we can assess whether the on-frame actor contributes to compositional aesthetics according to the rule-of-thirds. Focusing on actor-camera synchronization, here we refine scenario factors by considering only the actor pose and shot side (i.e., from which side the actor is shot by the camera) to categorize common cinematographic cases.

\begin{figure}[htbp]
\includegraphics[width=.5\textwidth]{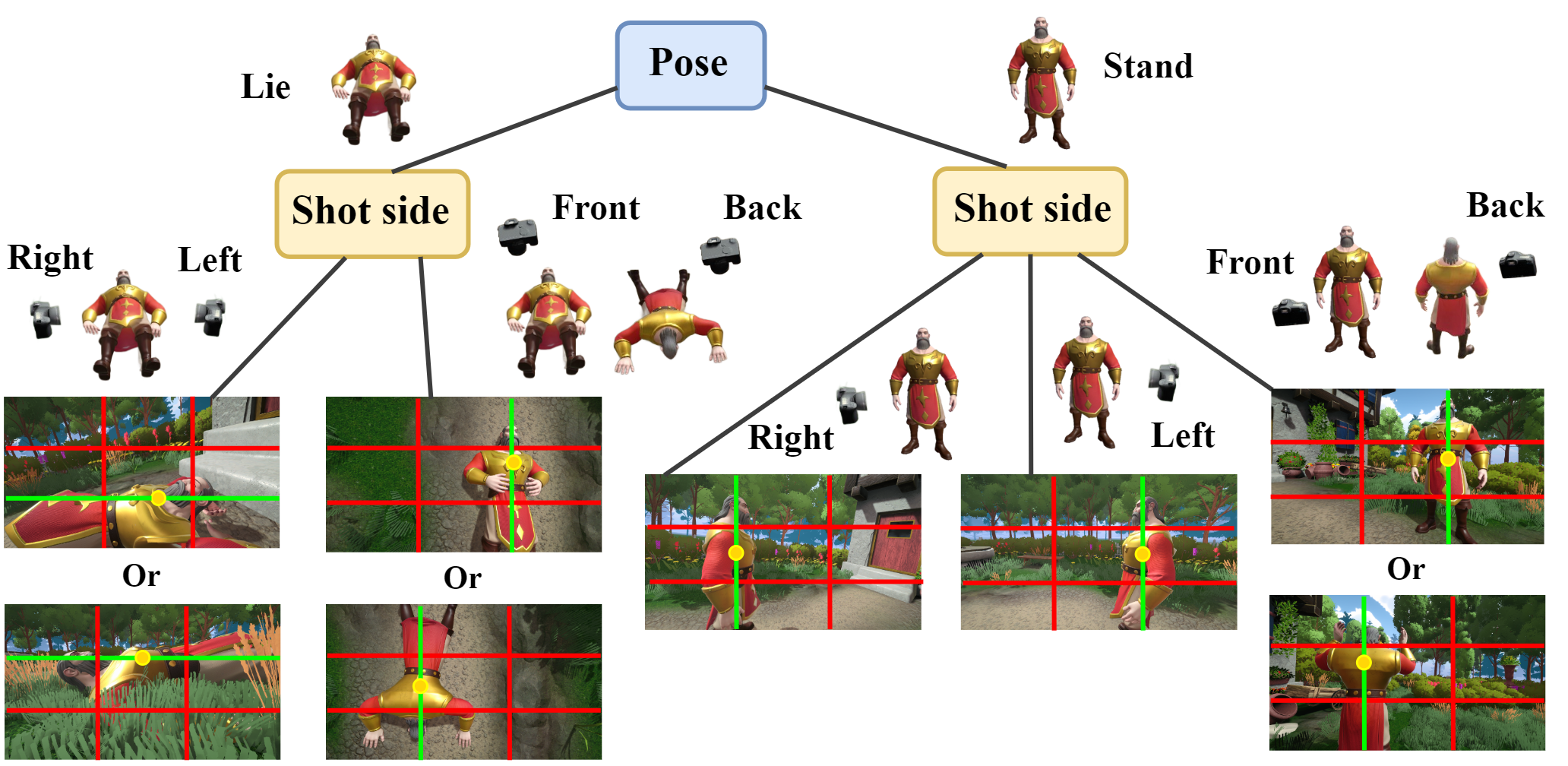}
\caption{Our rule-of-thirds decision tree. Based on different situations, the on-frame body center of the actor (marked as yellow dot) should stay on a certain alignment line (marked in green) to achieve compositional aesthetics.}
\label{RoTree}
\end{figure}

Fig. \ref{RoTree} illustrates a decision tree that specifies the aesthetic regulations utilized in our method. We divide each frame into a grid of four lines to mark one-third alignments. The actor's body, simplified to a weighted mean of joints projected onto the frame, should closely approach a certain alignment line under different cases for aesthetic framing. This transforms our adjustment into a problem discussing the distance between the projected body center and the alignment line. Building on this idea, we train a render-free adjustment network $\psi$ to understand camera projection during the modification process with aesthetic regularizers for self-supervision.

\begin{figure}[htbp]
\centering
\includegraphics[width=.45\textwidth]{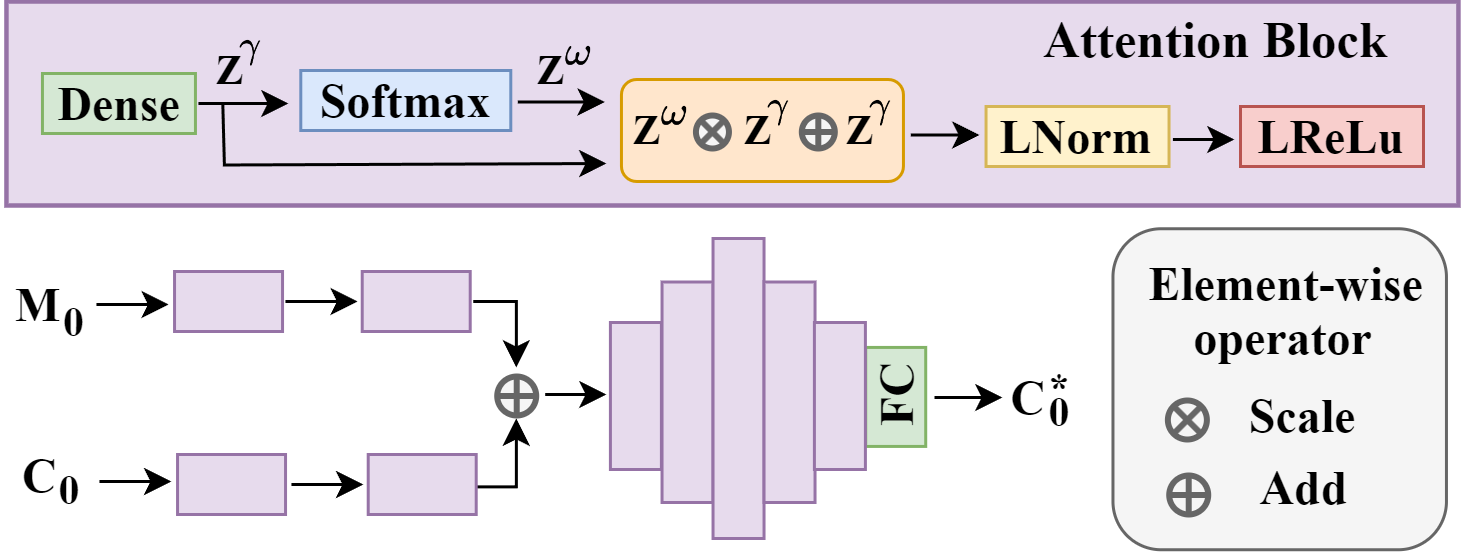}
\caption{The architecture of our adjustment network $\psi$.}
\label{AdjNet}
\end{figure}

\subsubsection{Network and self-supervised losses}
The network architecture of $\psi$ incorporates attention blocks \cite{vaswani2017attention}, leveraging their ability to weight features with high importance. These blocks are utilized to analyze initial actor pose $M_0$ and camera placement $C_0$, aiming to emphasize on-frame actor joints and specific camera axes that significantly influence shot composition, respectively. As depicted in Fig. \ref{AdjNet}, our model $\psi$ uses a dense layer to extract latent features $Z^{\gamma}$, which are then merged with self-attention features $Z^{\omega}$ through a scale-add operation, followed by layer normalization and a leaky ReLU activation. The captured camera and actor features are then fused together to imitate camera projection, enabling the estimation of an adjusted camera placement $C_0^{*}$ that facilitates immersive frame aesthetics.

The training of $\psi$ is self-supervised using a hybrid loss function $\mathcal{L}_{aes}$, which not only regularizes aesthetic frame composition but also ensures minimal adjustment in order to maintain user preference.

\vspace{1mm} 

\textbf{Composition loss:} Given $M_0$ and $C_0^{*}$, we use camera projection to calculate the projected 2D actor pose $M^{*}_{p} \in \mathbb{R}^{J \times 2}$. The on-frame body center is thus located by computing the weighted mean joint of $M^{*}_{p}$, which is denoted as $\overline{M^{*}_{p}} \in \mathbb{R}^{2}$. Using the decision tree shown in Fig. \ref{RoTree}, we obtain candidates of alignment lines as $A_l \in \mathbb{R}^{2 \times 2}$. The composition-driven rule-of-thirds constraint can be effectively formed via the point-to-line distance as:

\begin{equation}\label{rotloss}
\mathcal{L}_{cmp}=\min (||\overline{M^{*}_{p}}-A_{l}^{0}||_{2},||\overline{M^{*}_{p}}-A_{l}^{1}||_{2}),
\end{equation}
where $A_{l}^{0}$ and $A_{l}^{1}$ represent the two possible alignment candidates that are both $\in \mathbb{R}^{2}$. If only one alignment line is determined as candidate based on the decision tree, $A_{l}^{0}$ and $A_{l}^{1}$ are set equal. 

\vspace{1mm}
\textbf{Adjustment loss:} To prevent excessive modification over the inputted $C_0$, we constrain the extent of adjustment as:

\begin{equation}\label{adjloss}
\mathcal{L}_{adj}=||{C_0}-C_0^{*}||_2.
\end{equation}

\textbf{Visualization loss:} We make efforts to preserve the original shot type (e.g. full, medium, close shot) in $C_0$ to further avoid over-adjustment at the visualization level. Hence, a contrastive regularizer is designed to monitor the consistency of the actor's on-frame and off-frame joints before and after modification as:

\begin{equation}\label{visloss}
\mathcal{L}_{vis}=M_{b}^{*} \cdot (1-M_{b}) + (1-M_{b}^{*}) \cdot M_{b},
\end{equation}

where $\cdot$ denotes the dot product. $M_{b}^{*}$ and $M_{b}$ are both binary vectors $ \in \mathbb{R}^{J}$ and each element represents whether the corresponding joint is visible in the shot. These vectors are obtained by binarizing the 2D actor pose projections under the camera placement ${C}_0^{*}$ and ${C}_0$, respectively.

In summary, the complete aesthetic loss function $\mathcal{L}_{aes}$ for optimizing $\psi$ can be formulated as:

\begin{equation}\label{lossphi}
\mathcal{L}_{aes}=\lambda_{cmp}\mathcal{L}_{cmp}+\lambda_{adj}\mathcal{L}_{adj}+\lambda_{vis}\mathcal{L}_{vis},
\end{equation}
where each $\lambda$ denotes the corresponding weight for a certain loss component.

The derivation of all variables mentioned above is detailed in supplementary materials. By conducting such compositional adjustment, we obtain an aesthetic initialization of camera placement for the subsequent camera trajectory synthesis.

\subsection{Camera trajectory synthesis}

In this module, we generate camera trajectories that provide precise spatial and emotional synchronization with the actor to enhance immersive viewer experiences. Leveraging the power of adversarial learning, our GAN-based generative model extracts features from actor-space $X^{h}$, including both physical movements and psychological states preprocessed from $M$, $E$, and $C_0^{*}$. These features are then transformed into the final camera trajectory $C'$, which effectively learns tracking and styling techniques that mimic those of a human director in cinematic production. We specify $X^{h}$ as follows:

\begin{equation}
X^{h}=\{M, M^{v}, M^{d}, E, C^{f}\}, 
\end{equation}
where 

\begin{itemize}

	\item $M$, $M^{v}$, and $M^{d}$ are all $\in \mathbb{R}^{T \times JQ}$. $M$ represents joint locations per frame, whereas $M^{v}$ and $M^{d}$ denote the joint velocity and its absolute, respectively. Random noise is padded to $M^{d}$ and $M^{v}$ for temporal alignment with $M$.

\vspace{1mm}
	
	\item E $\in$ (0, $E_{max}$] and $E_{max}>1$. Following \cite{wu2018thinking}, we categorize actor emotion into tense or relaxed, where $E<1$ indicates a relaxed emotion, with lower values for greater relaxation, while $E>1$ denotes tension, the higher the tenser the emotion.

\vspace{1mm}
 
	\item  $C^{f} \in \mathbb{R}^{T \times Q}$ is obtained by repeating $T$ times of $C_{0}^{*}$ in the temporal domain, capturing initial correlations between the focused actor and the camera.

\end{itemize}

\vspace{1mm}
\subsubsection{Network architecture}

To improve efficiency in actor-to-camera processing, our generator $G$ utilizes an encoder-decoder architecture with intermediate latent representation to facilitate cross-space transformation.

\begin{figure}[htbp]
\centering
\includegraphics[width=.45\textwidth]{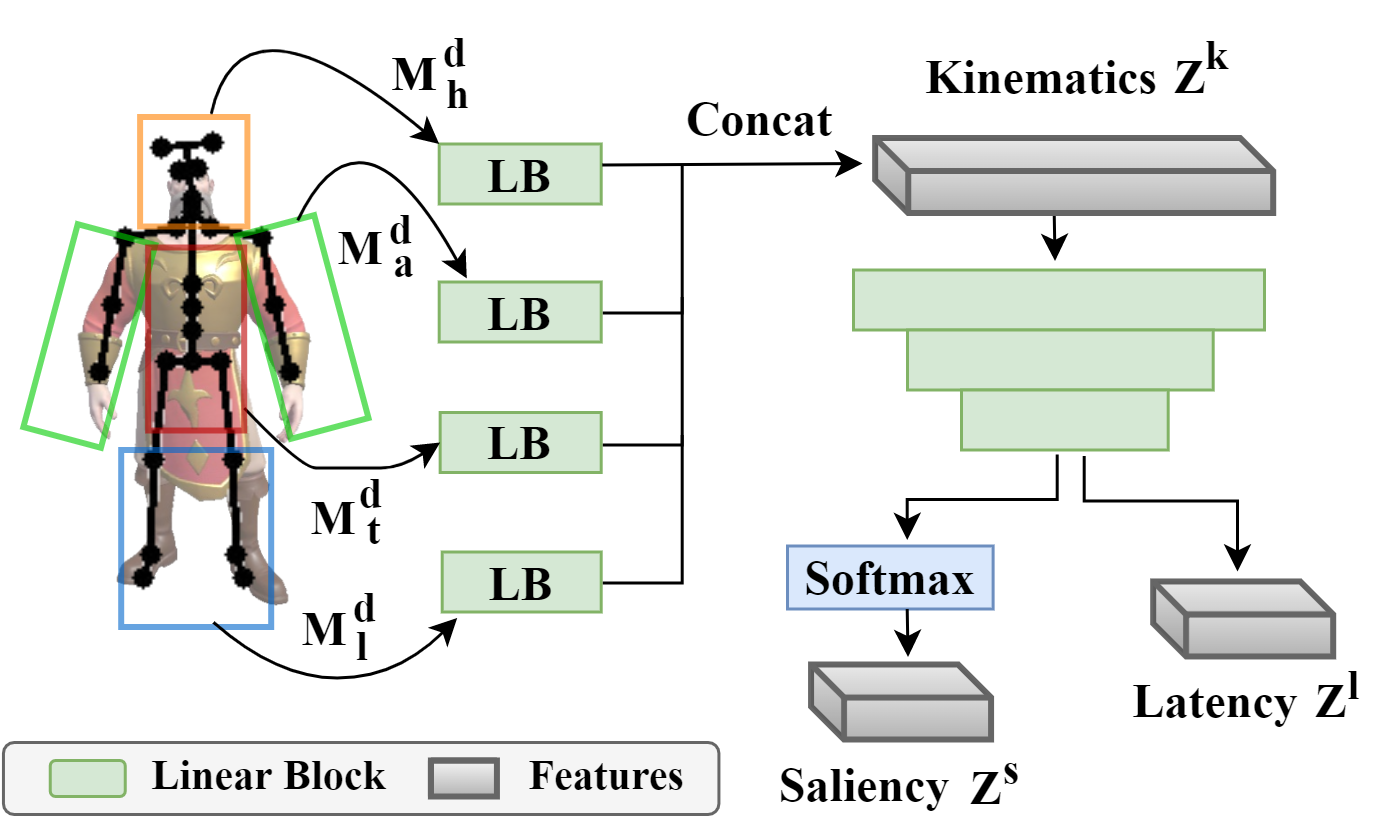}
\caption{The design of encoder in $G$. See text descriptions for details.}

\label{encoder}
\end{figure}

\textbf{Actor encoder:} During the encoding phase, actor poses are disarticulated into four parts: head, arms, torso, and legs, to enable a fine-grained analysis of kinematic features at low cost for precise spatial tracking. As illustrated in Fig. \ref{encoder}, the encoder takes $M^{d}$ as input, which is divided into $M^{d}_{h}$, $M^{d}_{a}$, $M^{d}_{t}$, and $M^{d}_{l}$, corresponding to different body regions. We leverage several linear blocks, each comprising a dense layer, a layer normalization, and a leaky ReLu activation, to extract region-wise kinematics. These local motion features are concatenated to form the overall kinematic embeddings $Z^{k}$. We then compress $Z^{k}$ through a couple of linear blocks to generate the latent feature $Z^{l}$. Additionally, a softmax operator is applied to derive a saliency map $Z^{s}$ that implicitly learns the hidden director tracking strategy.

\begin{figure}[htbp]
\centering
\includegraphics[width=.45\textwidth]{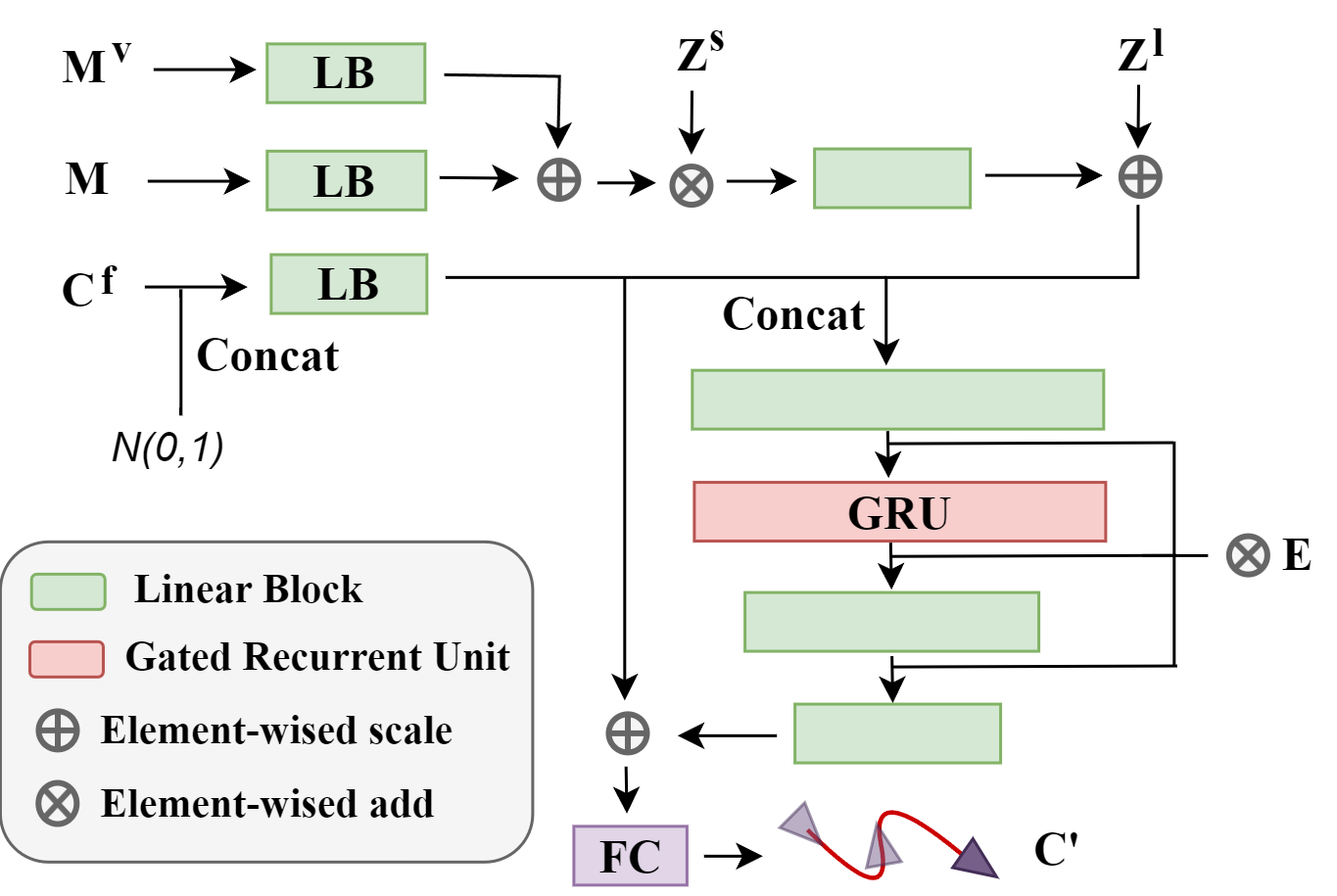}
\caption{The design of decoder in $G$. See text descriptions for details.}

\label{decoder}
\end{figure}

\vspace{1mm}
\textbf{Camera decoder:} The decoder additionally incorporates $M^{v}$, $M$, and $C^{f}$ to further strengthen the actor-camera synchronization during feature transformation, with random noise merged into $C^{f}$ to enhance the model's robustness. The encoded tracking strategy $Z^{s}$ and actor kinematics $Z^{l}$ are integrated into the latent decoding space via element-wise multiplication and addition, respectively. To effectively synthesize camera sequences, Gated Recurrent Units (GRUs) \cite{chung2014empirical}, a refined variant of recurrent neural networks (RNNs), are employed due to their superior capacity for handling sequential data. Moreover, the variable $E$, which indicates the user-desired emotional category and intensity, is repeatedly introduced in the generation. This enables appropriate emotional styling by adjusting the overall amplitude of the generated camera trajectory so as to match a certain actor emotion. Finally, a skip connection of features obtained from $C^{f}$ is applied at the end of the decoder to benefit efficiency, allowing the network to focus on learning only the temporal evolution when estimating the output $C'$.

\vspace{1mm}

\textbf{Discriminator:} Our discriminator $D$ adopts a Siamese \cite{chopra2005learning} architecture, which is powerful in distinguishing between real and fake samples, particularly for sequence-based data \cite{lee2019dancing}. It compares the generated $C'$ against ground-truth $C$ through two identical shared-weight branches, each consisting of three-layer linear blocks to capture cinematic dynamics. All the features extracted from the paired samples are then concatenated and fed into a classifier to estimate the similarity, where a higher similarity score denotes greater generation accuracy. This discriminator $D$ facilitates supervising our generator $G$ in an adversarial manner.

\vspace{1mm}

\subsubsection{Loss functions}\label{gloss} Our trajectory generator $G$ is trained using the adversarial loss $\mathcal{L}_{adv}$ and trajectory loss $\mathcal{L}_{tj}$. The latter one comprises three regularizers at point-level, shape-level, and feature-level compared to real professional samples.

\vspace{1mm}
\textbf{Point loss:} We calculate L2 distance between the ground-truth $C$ and our generated $C'$ with total variation \cite{mahendran2015understanding} for temporal smoothness as:

\begin{equation}\label{mse}
\mathcal{L}_{mse}=||C’-C||_{2}^{2}+\sum_{t=1}^{T-2} (||C'_{t+1}-C'_{t}||+||C'_{t}-C'_{t-1}||),
\end{equation}
where $T$ denotes the total number of frame time and $t$ ranging from 0 to $T-1$.

\vspace{1mm}
 \textbf{Shape loss:} Following \cite{emocam}, we control the amplitude of the camera trajectory to express varied psychological states of the actor. For precise emotional styling in the generated $C'$, we align the amplitude-based trajectory shape with that of the well-designed real sample $C$.

 \begin{figure}[htbp]
 \centering
\includegraphics[width=.45\textwidth]{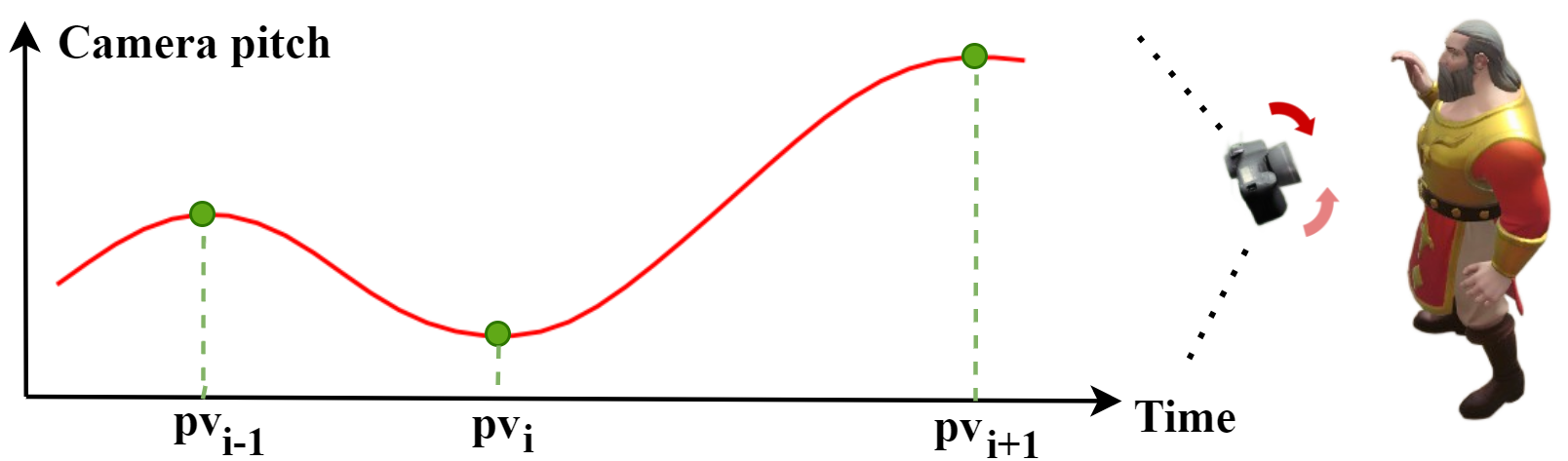}
\caption{An example visualizing the calculation of amplitude in the pitch axis. See text below for the detailed algorithm.}

\label{demoshk}
\end{figure}

As shown in Fig. \ref{demoshk}, we collect the time points of peaks and valleys along a specific camera axis, denoted as $Pv=\{pv_{0},pv_{1},...,pv_{N}\}$. The axis-wise amplitude can be analytically measured using $f_{amp}$ as:

\begin{equation}\label{shk}
f_{amp}(C^{q})= \sum_{i=1}^{N} \frac{|C_{{pv}_i}^{q}-C_{{pv}_{i-1}}^{q}|}{pv_{i}-pv_{i-1}},
\end{equation}

where $C^{q}$ denotes the trajectory in axis $q$ among $Q$ degrees of freedom. We use $i$ to iterate through the time point collected in $Pv$. The overall amplitude of $C$ can then be obtained by performing $f_{amp}$ on all the camera axes, named $C_{amp} \in \mathbb{R}^{Q}$. This allows us to regularize the trajectory shape generation as:

\begin{equation}\label{lossshk}
\mathcal{L}_{shp}=||C’_{amp}-C_{amp}||.
\end{equation}

\textbf{Feature loss:} Given the effectiveness of VGG \cite{simonyan2014very} in extracting perception-senstive features, we introduce it here to supervise the similarity between the real $C$ and our generated $C'$ in the feature space, as:

\begin{equation}\label{feat}
\mathcal{L}_{feat}=||VGG(C’)-VGG(C)||_{2}^{2}. 
\end{equation}

\textbf{Adversarial loss:} Taking the advantage of the GAN framework, our discriminator $D$ tries its best to distinguish between real and fake pairs through maximizing the loss:

\begin{equation}\label{disd}
\mathcal{L}_{adv}^{d}=\mathbb{E}[log{D(C,C)}]+\mathbb{E}[log(1-D(C’,C))].
\end{equation}

Based on the adversarial training, our generator $G$ is forced to improve the synthesized results in order to fool $D$ by minimizing the function:

\begin{equation}\label{disg}
\mathcal{L}_{adv}^{g}=\mathbb{E}[-log(D(C’,C))].
\end{equation}

In summary, combining all the regularizers above, the final loss function for the generator $G$ can be formulated as:

\begin{equation}\label{lossg}
\mathcal{L}_{g}=\lambda_{mse}\mathcal{L}_{mse}+\lambda_{feat}\mathcal{L}_{feat}+\lambda_{shp}\mathcal{L}_{shp}+\lambda_{adv}\mathcal{L}_{adv}^{g},
\end{equation}

where each $\lambda$ denotes the corresponding weight for a certain loss component.

Using the aesthetic adjustor $\psi$ and trajectory generator $G$, we can ensure that our camera control system outputs camera movements that satisfy a comprehensive actor-camera synchronization in terms of aesthetics, action, and emotion for significantly enhancing immersive viewer experiences.

\section{Experiments}

In this section, we provide training details of our camera control system and assess its generation accuracy via ablation studies. Then, we describe quantitative experiments as well as a qualitative user study that evaluates the viewing quality of cinematic videos produced by our system, focusing on immersion in spatial, emotional, and aesthetic aspects.

\subsection{Experimental environment}

\begin{figure}[htbp]
\centering
\includegraphics[width=.4\textwidth]{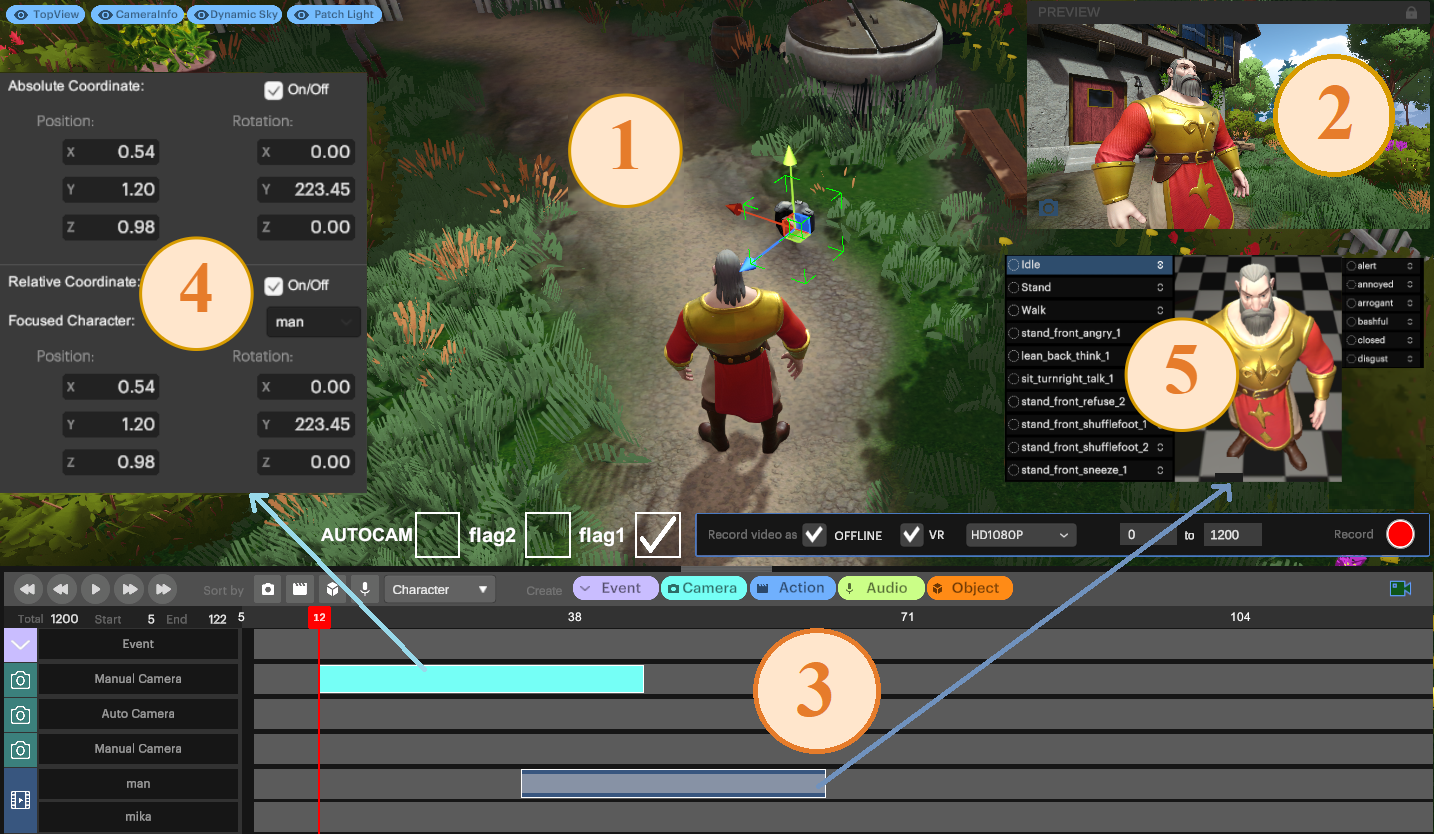}
\caption{Main functional zones in the environment. Zone 1 is used for the overall scene view and acting display, while Zone 2 visualizes the camera view. Zone 3 allows users to manipulate camera and character behaviors by dragging and dropping blocks on the timeline. Zone 4 and Zone 5 offers  detailed management for camera and character, respectively, via pop-up menus. They control configuration like the focused character of the camera, character action, and emotion variable.}
\label{userInt}
\end{figure}

As shown in Fig. \ref{userInt}, a Unity3D application is developed to create a 3D virtual environment with cinematic resources including scenes, characters, and cameras. This application provides users with interfaces to customize their cinematic works and monitor shooting through information panels or a real-time camera view window. Especially, it supports the export and import of designed actor and camera behaviors for research purposes. Our camera control system is finally integrated as a plugin within the environment, enhancing the sense of immersion in user-generated cinematic videos. For more information, please refer to the supplementary materials.

\subsection{Dataset}

Due to different training strategies, we build separate datasets for the self-supervised aesthetic adjustor $\psi$ and the supervised GAN-based camera trajectory synthesis model $G$. 

\vspace{1mm}

\textbf{Samples for training $\psi$}: To facilitate the self-supervised training of $\psi$, we combine synthetic camera placement $C_0$ and actor pose $M_0$ to simulate diverse shooting scenarios for adjusting frame aesthetics. Apart from a few frequently used locations collected from professional artists, we use sphere meshes around the focused actor at various radial distances, yielding 481,536 potential camera placements. Furthermore, we retrieve 57 typical initial poses from action resources for pose input. By pairing these $C_0$ and $M_0$ with invalid shots filtered out, our aesthetic adjustment dataset comprises 13,066,689 samples in total, which is subsequently split into 80\% for training and 20\% for testing.

\vspace{1mm}

\textbf{Samples for training $G$}: In our virtual environment, we ask artists to design camera movements given several director screenplays. During production, they are required to use cinematographic techniques such as spatial tracking and emotional styling to achieve actor-camera synchronization for immersive viewer experiences. Each training pair includes scene settings—action sequences $M$, emotion variable $E$, and aesthetic initial camera placement $C_0^{*}$—combined with the artist-designed camera trajectory $C$ as ground truth, all exported via environment interface. Due to limitations of human labor, these manual pairs constitute 20\% of our dataset, with the rest synthesized based on artist samples through offsetting, sequence flipping, and emotion-driven amplitude modification to increase data diversity. This results in a total of 25,230 five-second footage samples, involving 583 character actions and 5 typical emotion variables $\{0.5, 0.75, 1, 1.5, 2\}$, semantically indicating "relax-more," "relax-less," "neutral," "tense-less," and "tense-more," respectively. In the testing phase, we randomly select 5550 samples excluded from training to evaluate the performance of our system.

\subsection{Training details}

We implemented all the networks and losses in PyTorch, with some of our loss functions mathematically smoothed to ensure differentiability. Our adjustment network $\psi$ reaches convergence after 35 epochs of training, using a batch size of 1024 as well as the Adam optimizer \cite{kingma2014adam}. We initialize the learning rate with 0.002 and decrease it by 10x every 10 epochs. The weights for the loss component in $\mathcal{L}_{aes}$ are set as: $\lambda_{cmp}$=1, $\lambda_{adj}$=0.25, $\lambda_{vis}$=0.01, to maximize aesthetic compositions while preserving user input. 

For the GAN model, we pretrain the generator $G$ for 100 epochs using only $\mathcal{L}_{mse}$ with the Adam optimizer and a batch size of 10. The initial learning rate is set to 0.005 and gets decreased by a factor of 10 every 25 epochs to obtain convergence. Then, following the same optimizer and batch size, we train $G$ and the discriminator $D$ adversarially for 45 epochs, with both learning rates set to 0.0001. During this process, the loss components of $\mathcal{L}_{g}$ are weighted as follows: $\lambda_{mse}$=10, $\lambda_{feat}$=0.5, $\lambda_{shp}$=1, $\lambda_{adv}$=0.2, enabling a balanced trajectory accuracy across different levels.

Note here that all the $\lambda$ values mentioned above are determined experimentally based on optimal model performance.

\subsection{Ablation study}

In order to demonstrate the effectiveness of our aesthetic adjustor $\psi$ and camera trajectory generator $G$, we compare them with variant models based on our testing dataset across multiple metrics.

\vspace{1mm}
\subsubsection{Evaluation of aesthetic composition adjustment}

We evaluate the performance of our aesthetic adjustor $\psi$ by comparing it with its variants trained on different loss components. Our adjustments are assessed by rule-of-thirds shift (RoTSft), adjustment distance (AdjDis), and visibility accuracy (VisAcc). RoTSft offers a direct aesthetic evaluation by computing the on-frame distance between the actor body center and the one-third alignment lines. Additionally, we consider practical adjustment factors crucial to user preference by AdjDis and VisAcc. The AdjDis denotes the total adjustment distance calculated via mean absolute error (MAE), while VisAcc measures the percentage of body joints that are accurately visualized compared to their original visibility. For computational details please refer to Equation (\ref{rotloss}), (\ref{adjloss}), (\ref{visloss}).

\begin{table}[htbp]
\centering
\caption{Comparison of aesthetic adjustment performance}
\begin{tabular}{ c |c |c |c  } 
 \hline
 Model Name & RoTSft (px) $\downarrow$ & AdjDis (m) $\downarrow$ & VisAcc (\%) $\uparrow$\\
 \hline
 Original & 412.4 & n/a & n/a \\
$\psi$ w/o $\mathcal{L}_{adj}$~\&~$\mathcal{L}_{vis}$ & \textbf{97.8} & 0.7860 & 46.11\% \\ 
$\psi$ w/o $\mathcal{L}_{vis}$ & 116.0 & \textbf{0.1191} & 47.87\% \\
$\psi$ & 129.5  &  0.1473 & \textbf{71.08\%}\\

 \hline
\end{tabular}
\label{tab_phi}
\end{table}

As demonstrated in Table \ref{tab_phi},  our $\psi$ model achieves a good balance across all metrics. In comparison with the leading competitor for RoTSft and AdjDis, our model exhibits a minor shortfall by around 32 pixels shifting from the alignment line in 1080P-resolution frames  and about a 3cm of camera adjustment in the virtual environment. These differences are practically negligible. Notably, our $\psi$ model outperforms others in VisAcc, which significantly affects human perception and determines shot type (e.g. full, medium, close-up), with exceeding 23\% of the body joints. This indicates that our aesthetic adjustor $\psi$ effectively modifies camera placement aligning with rule-of-thirds aesthetics while optimally maintaining the original user-preferred shot designs.

\begin{figure}[htbp]
\centerline{\includegraphics[width=.45\textwidth]{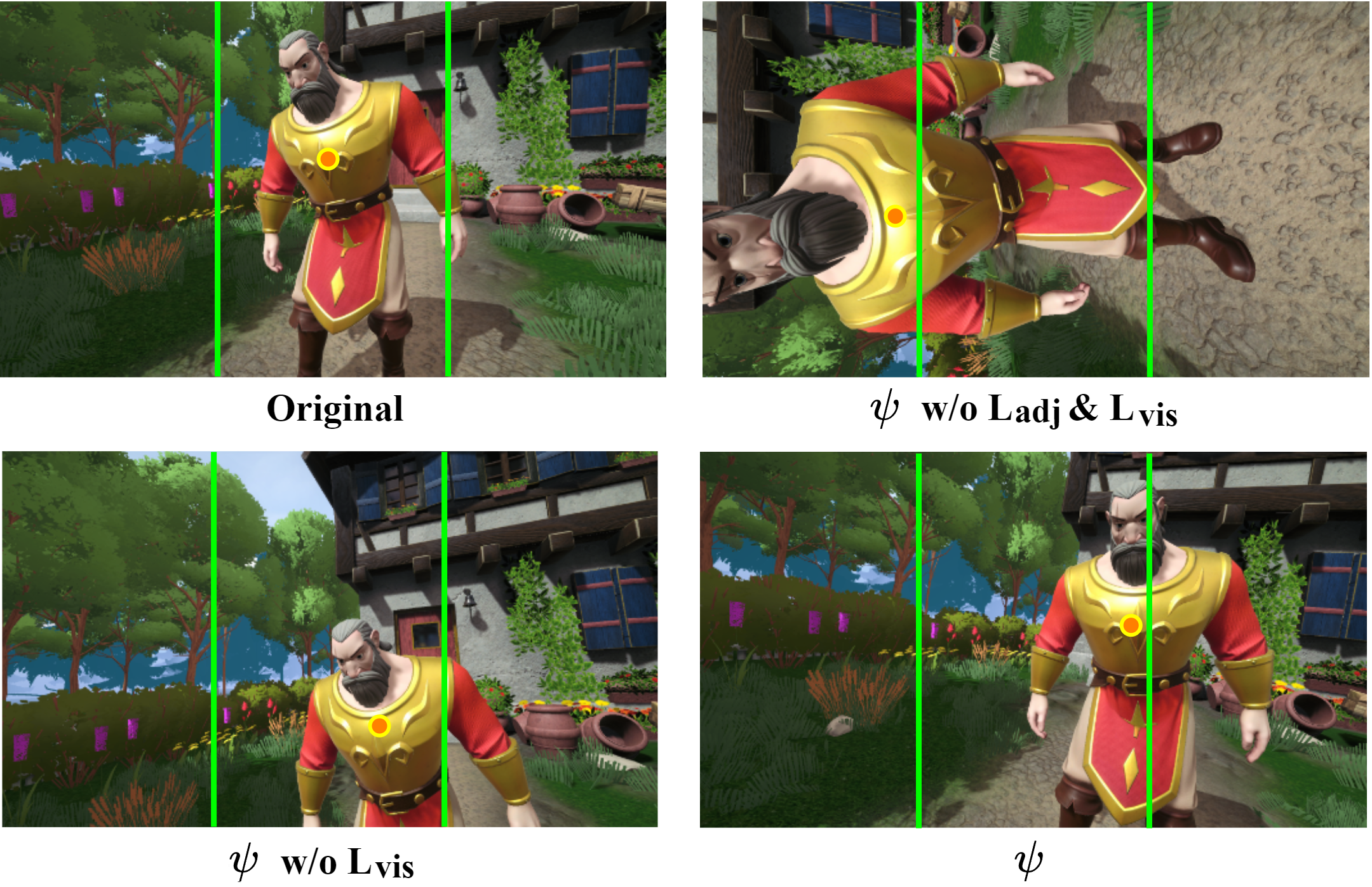}}
\caption{Example of qualitative comparison over aesthetic adjustment performance. The rule-of-thirds alignment candidates are labeled in green, whereas the orange dot represents the on-frame body center of the actor.}
\label{abl-rot}
\end{figure}

The qualitative example in Fig. \ref{abl-rot} illustrates the superior aesthetic adjustment capability of our $\psi$ model. Unlike the other two variant models, it prevents over-adjustment that arises from overfitting the rule-of-thirds compositional constraint or making excessive changes for actor joint visibility.

\vspace{1mm}
\subsubsection{Evaluation of camera trajectory synthesis} We compare our generator $G$ against its $\mathcal{L}_{mse}$-based pretraining model and two other GAN variants with partial loss components. Metrics including MSE, $\text{CosD}_{\text{A}}$, LPIPS \cite{simonyan2014very}, and FID \cite{heusel2017gans} are used to evaluate the models, where $\text{CosD}_{\text{A}}$ calculates the cosine distance between the overall amplitudes of real and synthesized trajectories, $C_{amp}$ and $C'_{amp}$, respectively (please refer to Equation (\ref{shk})). This enables us to comprehensively assess the generated camera trajectories based on ground-truth samples, considering point accuracy, shape consistency, and feature space similarity, from low to high levels.

\begin{table}[htbp]
\centering
\caption{Comparison of camera trajectory synthesis}
\begin{tabular}{ c |c |c |c |c } 
 \hline
 Model Name & MSE $\downarrow$ & $\text{CosD}_{\text{A}}$ $\downarrow$ & LPIPS $\downarrow$ & FID $\downarrow$\\
 \hline
Pretrain & \textbf{0.0082} & 0.2935 & 0.0510 & 0.0702\\
$G$ w/o $\mathcal{L}_{feat}~\&~\mathcal{L}_{shp}$  & 0.0149 & 0.2569 & 0.0477 & 0.0739\\
$G$ w/o $\mathcal{L}_{shp}$  & 0.0135 & \textit{0.2509} & \textbf{0.0385} & \textit{0.0681}\\
$G$  & \textit{0.0124} & \textbf{0.2380} & \textit{0.0405} & \textbf{0.0634}\\
 \hline
\end{tabular}

\label{tab_g}
\end{table}

In Table \ref{tab_g}, the best results are marked in bold, while the second-best are in italics. It reveals that the pretraining model outperforms in MSE but falls behind for the other medium-to-high level metrics, indicating its limitation in generating fine-detailed camera trajectories. The performance of the two GAN variants highlights the benefits of adversarial training and $\mathcal{L}_{feat}$ for an enhanced feature-space representation. However, they perform less effectively in $\text{CosD}_{\text{A}}$ compared to our $G$ model, which integrates $\mathcal{L}_{shp}$ that highly influences the shape of the trajectory. Generally, our $G$ model achieves the best trading-off, leading in  $\text{CosD}_{\text{A}}$ and FID with a slight compromise in MSE and LPIPS.

\begin{figure}[htbp]
\includegraphics[width=.45\textwidth]{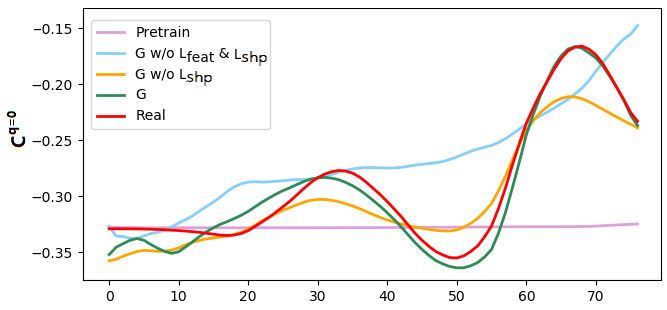}
\caption{A qualitative comparison of trajectories from different models, showing camera positions over the temporal evolution.}
\label{abl-1}
\end{figure}

Meanwhile, the qualitative example shown in Fig. \ref{abl-1} further supports our conclusions drawn from the table analysis. By utilizing a hybrid loss function, our $G$ model can accurately capture high-level features and refine the generated camera trajectory to match the shape of the real artist sample, thereby enabling the learning of immersive shooting techniques.

\subsection{Quantitative evaluation of immersion}

We quantitatively assess the immersive performance of our camera control system $\Upsilon$, which combines both the $\psi$ and $G$ models. This evaluation, following actor-camera synchronization, specifically focuses on three aspects: spatial action, emotional status, and frame aesthetics, all significantly impacting immersive viewer experiences. We conduct separate experiments to target these immersion-critical factors in comparison with other feasible methods.

\vspace{1mm}
\subsubsection{Spatial immersion} To evaluate the immersion arising from spatial actor-camera synchronization, we measure the consistency between character action features $Z_{s}^{a}$ and camera movement features $Z_{s}^{c}$ to quantify the spatial tracking performance. Due to cross-platform difficulty, we implement two advanced auto-cinematography methods from Burelli et al. \cite{burelli2016game} and Yu et al. \cite{yu2023novel} within our environment for comparison. Burelli et al. \cite{burelli2016game} propose attaching the camera fixedly to a specific actor joint based on the desired shot type for efficient action tracking. Yu et al. \cite{yu2023novel} offer an optimization-based camera controller that follows the focused actor considering fidelity and aesthetics. We tested these models across 37 virtually-staged screenplays, designed with character action $M$, initial camera placement $C_0^{*}$, and neural emotion variable (i.e., $E$=1). The mean action velocity, serving as ground-truth actor features $Z_{s}^{a}$, and the corresponding camera velocity $Z_{s}^{c}$ are compared together using the Hausdorff Distance (HD) for calculating feature-level similarity. Both features are normalized to [0,1] and structured as $\mathbb{R}^{(T-1) \times Q}$.

\begin{figure}[htbp]
\includegraphics[width=.48\textwidth]{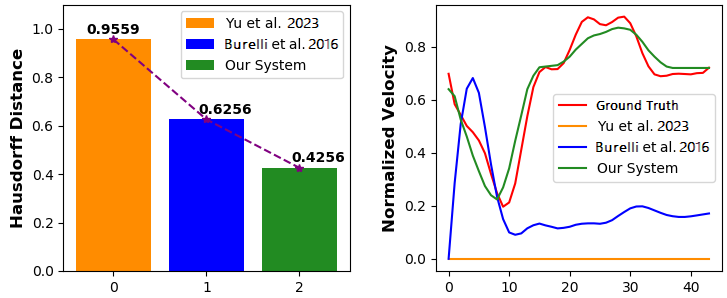}
\caption{Comparison of immersive performance based on spatial tracking accuracy. Left: HD results, the lower the better. Right: An qualitative example showing spatial actor-camera synchronization from different models over time.}
\label{spatial}
\end{figure}

The Hausdorff Distance (HD) results in the left of Fig. \ref{spatial} demonstrate that our system significantly surpasses the other two models in generating camera features consistent with the performed action. This improvement is attributed to our actor-to-camera generation framework. The qualitative example on the right side of Fig. \ref{spatial} further verifies that our method can smoothly track actor movements in real time, which outperforms the periodic updating strategy in Yu et al. \cite{yu2023novel}. Moreover, compared to Burelli et al. \cite{burelli2016game}, the use of saliency maps in our method avoids constant focus on a fixed body joint during camera trajectory synthesis, thereby better preserving the overall trend of character action.

\vspace{1mm}

\subsubsection{Emotional immersion} We assess the immersive performance of emotional actor-camera synchronization by analyzing the feature-based correlation between actor emotion status and camera behavior. Given the lack of existing methods addressing such an emotional immersion, we simulate two variant models based on our method as feasible competitors, named PlainCam and EmoCam. The PlainCam is obtained using our dataset with emotion-related variances removed, and for the EmoCam, we disable the use of $\mathcal{L}_{shp}$ in its training. The evaluation involves the same screenplays from the spatial immersion analysis, with each screenplay tested under 5 typical emotion variables $E \in \{0.5, 0.75, 1, 1.5, 2\}$, representing diverse psychological states of the actor. These emotion variables constitute a vector acting as the ground-truth emotional actor feature $Z^{a}_{e}$. Correspondingly, for each $E$, we average the trajectory amplitude $C’_{amp}$ and combine them into a vector as our camera stylistic feature $Z^{c}_{e}$. Hence, the immersion from emotional styling can be evaluated by calculating various correlation coefficients between $Z^{a}_{e}$ and $Z^{c}_{e}$, both of which are in $\mathbb{R}^{5}$.

\begin{table}[htbp]
\centering
\caption{Comparison of emotional immersion over behavior correlation }
\begin{tabular}{ c |c |c |c  } 
 \hline
 Model Name & PCC $\uparrow$ & SRCC $\uparrow$ & KRCC $\uparrow$ \\
 \hline
PlainCam & 0.7748 & 0.7822 & 0.7607 \\ 
EmoCam  & 0.8017 & 0.8112 & 0.7652 \\
Our System $\Upsilon$  & \textbf{0.9235} & \textbf{0.9356} & \textbf{0.9153} \\
 \hline
\end{tabular}

\label{tab_emoshk}
\end{table}

Table \ref{tab_emoshk} presents three correlation coefficients: Pearson (PCC), Spearman Rank (SRCC), and Kendall Rank (KRCC), to measure the relationship between actor emotion and our generated camera behavior. It is evident that, compared to other models, our method outperforms in synthesizing camera trajectories with adaptive amplitudes that closely align with the actor’s emotional states. This benefits from the emotion-related specifics incorporated into our network architecture and loss functions, enabling effective emotional actor-camera synchronization to achieve an enhanced sense of immersion

\begin{figure}[htbp]
\includegraphics[width=.48\textwidth]{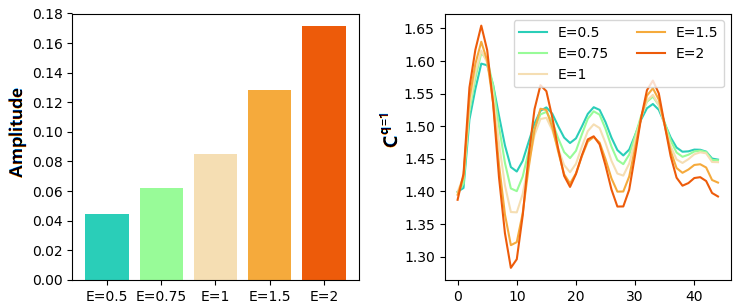}
\caption{Visualization of our generated adaptive camera trajectory amplitude. Left: Mean amplitude per sample over emotion variables. Right: A qualitative example showing camera trajectory under different $E$ over time.}
\label{emotional}
\end{figure}

Fig. \ref{emotional} provides an in-depth look at the emotional styling ability of our system. On the left, we can observe a transition from low to high trajectory amplitudes in response to varied emotion variables $E$, ranging from relaxation to tension. This imitates professional cinematographic techniques to enhance emotional immersion. On the right, despite the amplitude adjustments for different $E$ settings, the generated trajectories maintain consistent motion trends crucial for accurate spatial tracking. Note that although the experiment focuses on a set of pre-defined $E$ values, practically our camera control system is capable of processing any positive $E$ inputted by users.

\vspace{1mm}

\subsubsection{Aesthetic immersion} To verify the ability in achieving composition-based aesthetic immersion, we compare our method with the state-of-the-art camera controller Yu et al. \cite{yu2023novel} and two leading aesthetic cropping models Li et al. \cite{li2018a2} and Hong et al. \cite{hong2021composing}. These models are all known for their performance enabling photography-level aesthetic framing. Utilizing the Aesthetic Visual Analysis (AVA) model \cite{zhao2020representation}, we evaluate the aesthetic quality of the produced cinematic videos by analyzing compositional features from the given frames and predicting scores that reflect human aesthetics judgments. Both Yu et al. \cite{yu2023novel} and our system are tested across 24 randomly selected screenplays with varied 3D stage settings, yielding 216 frames evenly sampled from videos generated by each method for evaluation. Meanwhile, the image-based cropping models Li et al. \cite{li2018a2} and Hong et al. \cite{hong2021composing} are assessed based on processing their results using frames from Yu et al. \cite{yu2023novel}.

\begin{table}[htbp]
\centering
\caption{Comparison of aesthetic immersion over AVA score}
\begin{tabular}{ c |c |c |c |c |c  } 
 \hline
 Model \textbackslash\, Scene & Library & Forest & Park & Bedroom & Average \\
 \hline
Yu et al. \cite{yu2023novel} & 4.6017 & 4.8571 & 4.5665 & 4.2329 & 4.5645 \\ 
Li et al. \cite{li2018a2}  & 4.5512 & 4.8643 & 4.5070 & 4.2274 & 4.5375 \\
Hong et al. \cite{hong2021composing}  & 4.5645 & 4.8695 & 4.5185 & 4.2498 & 4.5506 \\
Our System $\Upsilon$  & \textbf{4.8077} & \textbf{5.0970} & \textbf{4.8556} & \textbf{4.4089} & \textbf{4.7923} \\
 \hline
\end{tabular}

\label{tab_aes}
\end{table}

\begin{figure*}[htbp]
\centerline{\includegraphics[width=.99\linewidth]{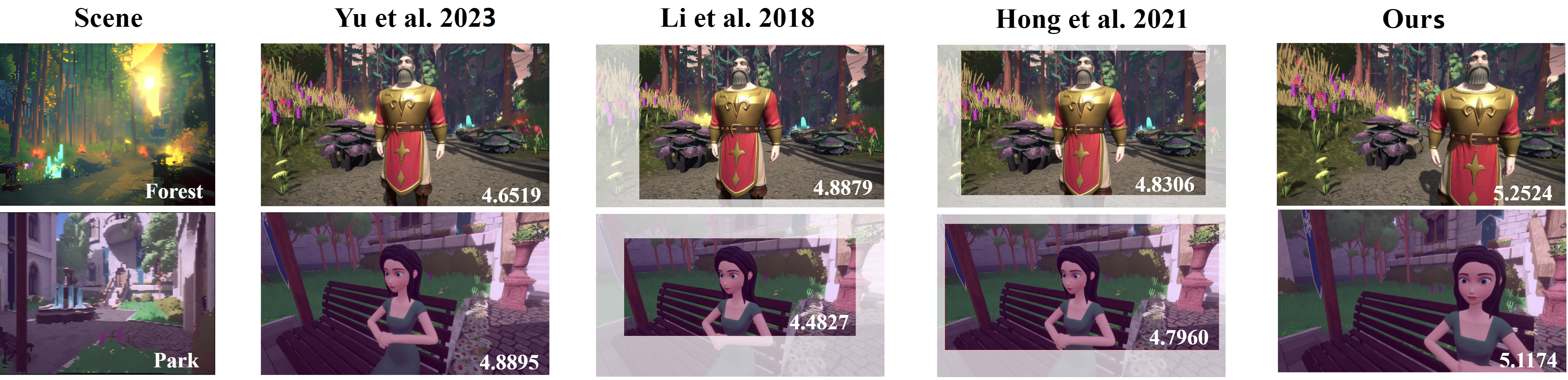}}
\caption{Qualitative examples for aesthetic comparison. We display frames from different models with the corresponding aesthetic scores predicted by \cite{zhao2020representation}.}
\label{aes_example}
\end{figure*}

Table \ref{tab_aes} shows the tested aesthetic scores for 4 typical indoor and outdoor scenes, where higher scores indicate better frame aesthetics. The two cropping models Li et al. \cite{li2018a2} and Hong et al. \cite{hong2021composing} can improve aesthetic frame composition beyond Yu et al. \cite{yu2023novel}, though limited to specific scenes. Conversely, our method robustly earns the highest aesthetic scores by incorporating the rule-of-thirds aesthetic principle. It’s worth noting that, although our aesthetic adjustments primarily focus on the initial camera placements, the use of spatial tracking ensures the preservation of these adjusted aesthetic compositions throughout the entire cinematic production, which effectively contributes to achieving aesthetic immersion. We also present examples in  Fig. \ref{aes_example} to qualitatively verify the aesthetic improvements of our method that strictly follows one-third alignments, compared to other models.

The quantitative experiments described above have separately demonstrated the effectiveness of our method in facilitating actor-camera synchronization through spatial tracking, emotional styling, and aesthetic framing, thereby jointly enhancing the overall immersive viewer experiences. In the next subsection, we conduct a user study to further evaluate the perceptual immersion of our generated cinematic videos in a qualitative way.

\subsection{Qualitative evaluation of immersion}\label{qualieval}

\begin{figure*}[htbp]
\centerline{\includegraphics[width=0.99\linewidth]{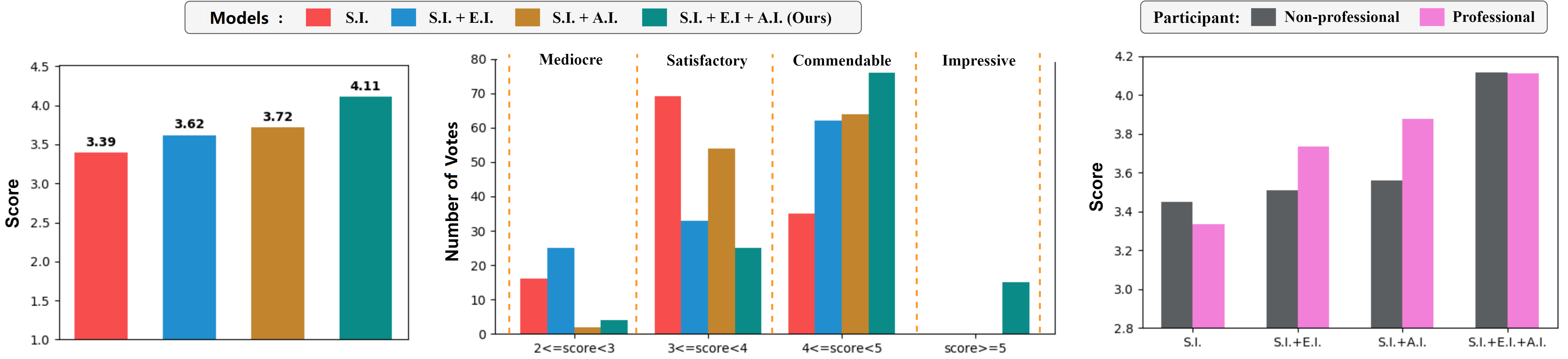}}
\caption{Rating results of immersive performance compared to the baseline Yu et al. \cite{yu2023novel}, with higher scores being better. Left: Average scores for different models. Middle: Histogram of score distributions. Right: Comparison of scores between professional and non-professional participants.}
\label{us_res}
\end{figure*}

A user study is additionally carried out to qualitatively assess the immersive performance of our camera control system. We compare our system against Yu et al. \cite{yu2023novel} and three variants of our methods, each emphasizing different actor-camera synchronization aspects: spatial (S.I.), spatial-emotional (S.I. + E.I.), and spatial-aesthetic (S.I. + A.I.) for distinctively creating immersion. Twelve participants are invited, half with cinematography knowledge while half without. Given screenplays, these participants play the role of directors, watching and blindly rating 10 sets of generated cinematic videos on a scale from 1 to 5. Yu et al. \cite{yu2023novel} serves as a baseline for the judgment, with scores $\geq 3$ indicating that our method or its variant models possess a greater capability to immersively convey the story. Conversely, a score below 3 means lesser effectiveness. The detailed procedures of this study are available in the supplementary materials.

Fig. \ref{us_res} demonstrates the user study outcomes. The left part shows that all four models averagely score above 3, outperforming the baseline Yu et al. \cite{yu2023novel}, with ratings increasing as the models integrate additional actor-camera synchronization techniques. This underscores the importance of each utilized shooting technique and the collective impact of our comprehensive method in enhancing immersion. The middle graph presents the detailed score distribution over the number of votes. Unlike other variant models, our method achieves leading stable viewer satisfaction by thoroughly addressing immersion across several perceptual dimensions. On the right, the comparison between professional and non-professional participants suggests professionals are relatively more sensitive to cinematographic changes, while both groups agree on the highest immersive enhancement provided by our model. 

\begin{figure*}[htbp]
\includegraphics[width=.96\textwidth]{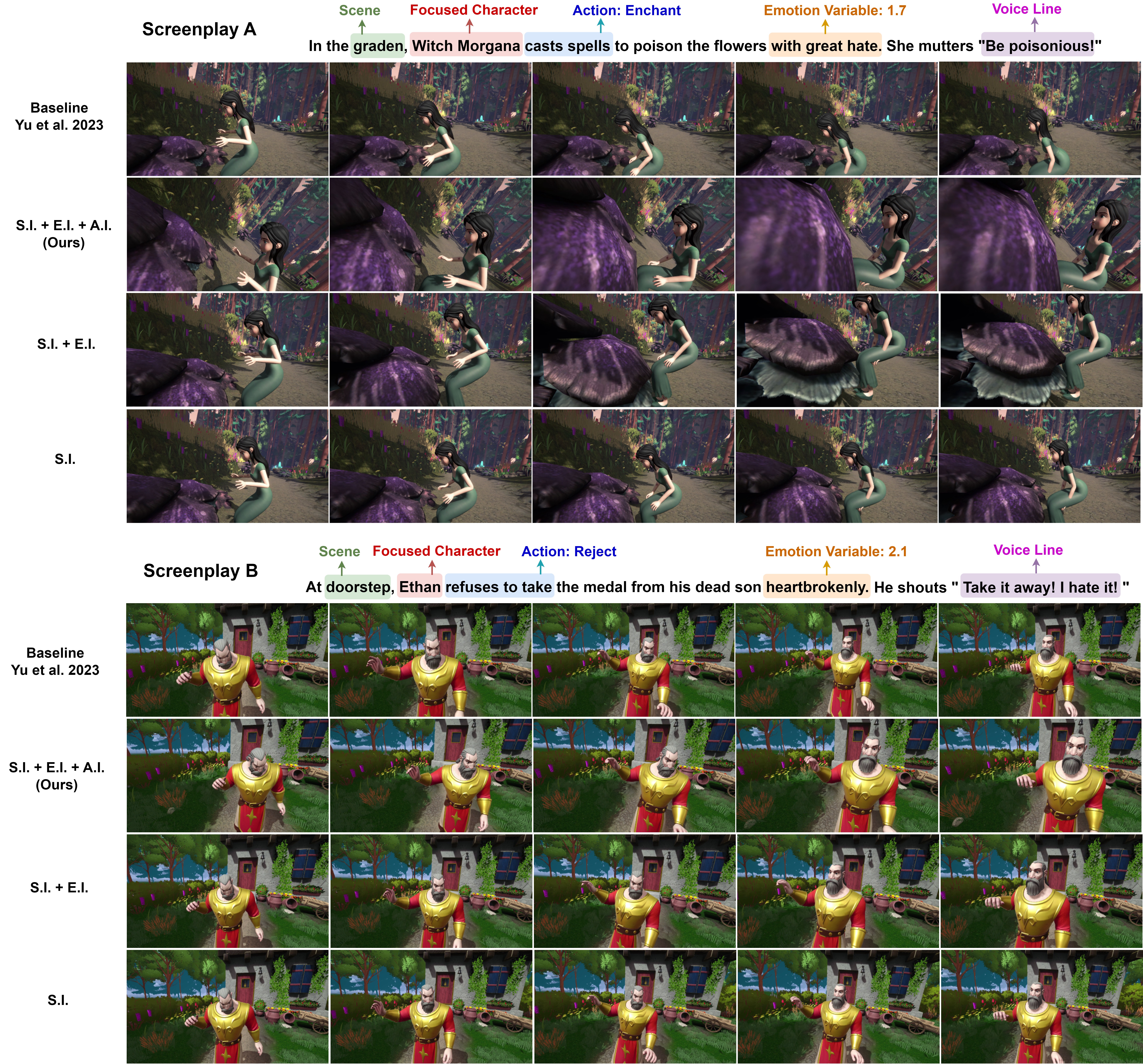}
\caption{Qualitative comparison of immersive performance across cinematic videos from different models. Frame sequences should be read from left to right.}
\label{qualiemp}
\end{figure*}

Fig. \ref{qualiemp} presents two cinematic examples utilized in our user study for qualitative comparison. Model S.I. provides a smooth and continuous spatial tracking of the focused character, which surpasses the baseline Yu et al. \cite{yu2023novel}. The addition of E.I. strengthens the amplitude of camera movement to more vividly express the desired character emotion. By incorporating A.I. for aesthetic control, our method further achieves a comprehensive balance among the three aspects of actor-camera synchronization techniques. This helps alleviate the potential out-of-frame issues (like those observed with Model S.I+E.I. in Screenplay A) and, all the used shooting techniques jointly contribute to the overall enhancement of immersive viewer experiences. For live demonstrations, please refer to the supplementary video.

\section{Discussion}

Our camera control system shows outstanding immersive enhancement compared to other methods. When deployed as a plugin in the 3D virtual environment, it allows users to adjust variable settings and repeatedly generate camera movements for obtaining ideal results. Our system is also adaptable to long-sequence and multi-person scenarios, treating them as independent footage segments according to the user-approved expression flow and shifting of the focused character. Due to establishing on well-recognized empirical rules, the diversity of cinematographic principles and personal preferences of users might affect the effectiveness of our method. Moving forward, we plan to extend our framework with a broader range of shooting techniques to enrich user selection. Additionally, considering the impact of other cinematographic factors like camera intrinsics and lighting on immersion, we aim to introduce more degrees of freedom, such as screen-based representations, to further enhance our toolkit and approach the capabilities of professional cinematic production.

\section{Conclusion}

In this paper, we propose a novel auto-cinematographic method for facilitating user-generated cinematic videos with enhanced sense of immersion. This is achieved by planning immersive camera movements following real-world cinematographic rules. More specifically, given the user-preferred setups, our camera control system synchronizes the camera with the focused actor across the aesthetic, spatial, and emotional levels. In the 3D stage, we design a deep camera control framework comprising an aesthetic adjustor and a camera trajectory synthesis model. The adjustor leverages the rule-of-thirds principle to conduct composition-based aesthetic framing self-supervisedly through camera projection analysis. Building on this aesthetic initialization, our GAN-based trajectory generator employs an encode-decoder architecture, mapping actor kinematics and emotional variables into camera movements. This ensures precise spatial tracking and emotional styling with constraints controlling trajectory accuracy and stylistic variances. The experiment demonstrates that our method outperforms other competitive models significantly in enhancing immersive viewer experiences under both quantitative and qualitative assessments.

\begin{figure*}[htbp]
\centerline{\includegraphics[width=.96\textwidth]{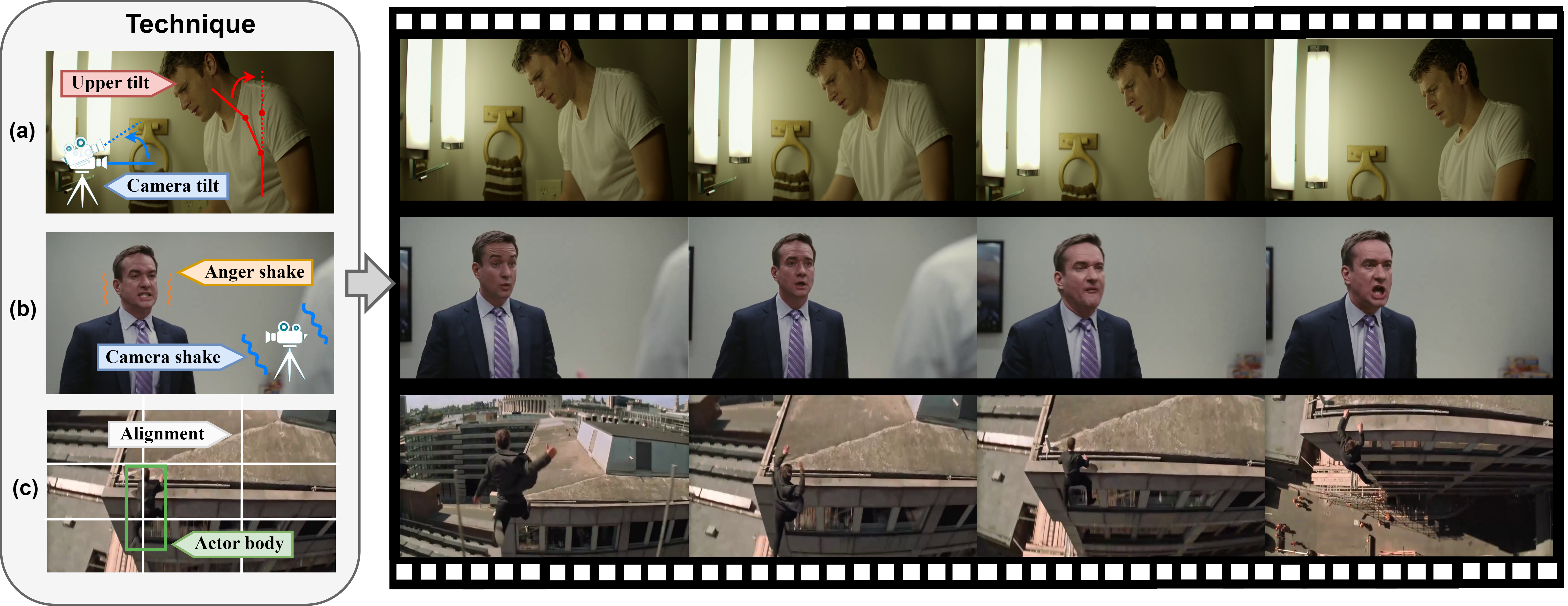}}
\caption{Real-world shooting techniques and their corresponding cinematic samples. (a), (b), and (c) exemplify how the actor-camera synchronization is addressed using camera movement at spatial, emotional, and aesthetic levels, respectively. See text above for detailed explanation. The live clips are available in the supplementary video.}
\label{FilmEmp}
\end{figure*}

\section{Supplementary Materials}

\subsection{Cinematographic knowledge}

In cinematography, directors strive to create an immersive experience so as to enhance viewer engagement and foster resonance between the actor and the audience. To produce high-quality immersive works, creators had to consider two fundamental questions: (1) What factors contribute to cinematographic immersion? and (2) How can these factors be effectively achieved?

Traditionally, immersion is often achieved by dynamic camera movements that simulate the perspective of an invisible character within the scene \cite{camproxy}. Directors typically focus on three key perceptual factors—spatial, emotional, and aesthetic—to enhance the immersive experience in practical productions \cite{bjork2005games,macklin2019going}. 
Building upon the well-recognized actor-camera synchronization principle \cite{hijackeye}, directors have broadened its applicability and adapted it to handle different perceptually critical aspects, which can be summarized as follows:

\begin{itemize}

 \item Synchronize camera with actor's physical movements for spatial-level immersion

\vspace{1mm}
 
 \item Synchronize camera with actor's mental state for emotional-level immersion

\vspace{1mm}
 
 \item Synchronize camera with pleasing on-frame locations of the actor for aesthetic-level immersion

\end{itemize}

Various shooting techniques are proposed to tackle these rules. Among them, the most general and straight-forward ones are spatial tracking, emotional styling and aesthetic framing. We detail each of them with samples demonstrated in Fig. \ref{FilmEmp}.

\vspace{2mm}

\textbf{Spatial tracking:} A widely used technique to achieve spatial-level immersion, which requires the camera closely following the actor's movements in a scene \cite{brown2016cinematography}. This involves capturing simple transition-based actions, such as running and walking, as well as complex fine-grained actor behaviors (which are primarily focused in this paper) like agree, refuse, curse, etc. We show an example of \textit{Mindhunter} directed by \textit{David Fincher} in Fig. \ref{FilmEmp}(a), where a camera tilt is employed to synchronize with the rise-up of the actor’s upper body, creating an observational view for the audience.

\vspace{2mm}

\textbf{Emotional styling:} A popular technique for creating emotional-level immersion involves styling the camera behavior to reflect the psychological state of the actor \cite{mekas1962note}. This styling can manifest as variations in the amplitude of camera movements—either intensified or weakened—to match the intensity and specific type of the actor's emotion, thereby conveying the corresponding mental feeling effectively to the audience \cite{emocam}. An illustrative example can be seen in Fig. \ref{FilmEmp}(b) from the series \textit{Succession}, directed by \textit{Mark Mylod}. In a scene where an actor exhibits extreme anger with trembling, Mylod utilizes exaggeratedly magnified camera shakiness to make the audience visually stroked and emotionally engaged with the actor.

\vspace{2mm}

\textbf{Aesthetic framing:} Achieving aesthetic-level immersion often needs control over frame composition according to aesthetic principles. For instance, following the rule-of-thirds principle \cite{rotcam}, the actor's body should be positioned at one-third of the frame throughout the shooting, taking into account compositional factors like the actor's pose and leading room \cite{may2020essential}. In Fig. \ref{FilmEmp}(c), an example is presented from \textit{Mission Impossible - Fallout} directed by \textit{Christopher McQuarrie}. In this scene, despite dynamic camera movements, the actor's body remains aligned with a certain third of the frame to create perceptually pleasing cinematic sequences for the audience's attention and affection.

\vspace{1mm}

In practice, a combination of these techniques is usually employed to optimize the immersive viewer experience. Beyond what has been discussed in this section, many other cinematographic rules and techniques, both camera-driven and not, can also contribute to the enhancement of immersion. While our paper mainly addresses the three crucial aspects of immersion using spatial, emotional, and aesthetic actor-camera synchronization, this foundational approach can be expanded with additional cinematographic modules to further benefit cinematic production comprehensively in the future.

\subsection{Implementation of aesthetic loss}
In this section, we provide implementation details related to camera projection, decision tree, and the derivation process of the variables used in our approach.

\IEEEpubidadjcol

\vspace{2mm}

\textbf{Camera projection:} To simplify the explanation, we denote the process of camera projection as $f_{p}$. Suppose there is an arbitrary pair of actor's single joint $P_j$ and camera placement $C_0$, where both of them are $\in \mathbb{R}^{Q=6}$ based on the 3D virtual environment. By performing $f_p(P_j, C_0)$, we can obtain the projection result $P_{2d} \in \mathbb{R}^{2}$. This allows us to map the coordinate from the 3D world to the 2D shot. 

The entire process can be divided into three main steps. Initially, the first three elements of $P_j$ indicating positions on the x,y,z axes, are extracted to construct the world coordinate $P_w=[X_w, Y_w, Z_w,1]^{T}$. This coordinate is then transformed into the camera coordinate $P_c=[X_c, Y_c, Z_c,1]^{T}$ via transformation matrix constructed according to the corresponding rotation and position of $C_0$. Subsequently, $P_c$ is projected onto the sensor plane $P_i=[x,y,1]^{T}$ based on the focal length of the camera. The final stage involves scaling and translating $P_i$ using intrinsic camera parameters to yield the shot coordinate $P_s=[u,v,1]^{T}$. The resulting 2D projection $P_{2d}=(u,v)$ can then be derived by extracting the corresponding values from $P_s$. Utilizing this process, we can project the entire actor pose $ M_0 \in \mathbb{R}^{J \times Q}$, given the camera placement $C_0$, to $ M_p \in \mathbb{R}^{J \times 2}$ through multi-dimensional matrix operations.

\vspace{2mm}

\textbf{Projection-related variables}: Based on the camera projection process $f_p$, the variables utilized in our loss components are derived as follows:

\begin{itemize}
	\item $M_{p}^{*} \in \mathbb{R}^{J \times 2}$ is obtained by $f_p(M_0, C_0^{*})$ using the aesthetically adjusted camera placement $C_0^{*}$. Based on the actual size of shot, joints detected off-frame are set to $(0,0)$ in the $M_{p}^{*}$.

 \vspace{1mm}
	
	\item $\overline{M_{p}^{*}}$ is obtained by computing the weighted mean of $M_{p}^{*}$, with the weights following a normal distribution. The torso center joint is assigned a weight corresponding to the peak of the distribution, while the weights of the other joints are decreased depending on their distances to the torso center. Note that off-frame joints are excluded from this operation. This ensures the body center is controlled to be robustly close to the torso whatever the visibility of the actor pose is.

  \vspace{1mm}
 
	\item $M_{b}^{*}$  is derived by binarizing $M_{p}^{*}$ to indicate whether a certain joint is on-frame (labeled as 1) or off-frame (labeled as 0). This is achieved by multiplying the projected $u$ and $v$ coordinates of each joint in  $M_{p}^{*}$ and using 0 as a threshold for binarization.

  \vspace{1mm}

    \item $M_{b}$ is obtained following the same process for $M_{b}^{*}$. Instead of using the adjusted camera placement $C_0^{*}$, its derivation is based on the initial camera placement $C_0$ to obtain the original shot preference from user input.
 
\end{itemize}

\vspace{2mm}

\textbf{From decision tree to $A_l$ }: Referring to the main text, we use a decision tree to determine the alignment lines $A_l$ that should be followed to improve aesthetic composition based on the rule-of-thirds principle. In this subsection, we provide detailed derivation process for obtaining $A_l$ under the consideration of actor pose and shot side. To indicate whether the actor is in a lying or standing pose, we compute the height difference between the actor's head and pelvis joints, which is denoted as $hpd$. Additionally, the side of the shot is represented by the relative angle $ra$ between the camera and the actor. Given that the on-frame body center $\overline{M_{p}^{*}}=(u_{m},v_{m})$ has been obtained and each shot is of size $(width, height)$, the candidates of alignment $A_l=\{(u_1, v_1), (u_2, v_2)\}$ can be derived as shown in Algorithm 1.

\begin{algorithm}
$\{(u_1,v_1),(u_2,v_2)\} \gets \{(0,0),(0,0)\}$\;
$thres \gets \text{Lie-to-stand threshold}$\;

\eIf(\tcp*[f]{Confirmed stand pose}){$hpd \geq thres$}
{
   \uIf(\tcp*[f]{Confirmed right shot}){$ra \in [45^{\circ},135^{\circ}]$}{
   $u_1,u_2 \gets \frac{1}{3}width$\;
   $v_1,v_2 \gets v_m$\;
  }
  \uElseIf(\tcp*[f]{Confirmed left shot}){$ra \in [225^{\circ},315^{\circ}]$}{
   $u_1,u_2 \gets \frac{2}{3}width$\;
   $v_1,v_2 \gets v_m$\;
  }
  \Else(\tcp*[f]{Confirmed front or back shot}){
     $u_1 \gets \frac{1}{3}width$\;
     $u_2 \gets \frac{2}{3}width$\;
     $v_1,v_2 \gets v_m$\;
  }
}{
    \eIf{$ra \in [45^{\circ},135^{\circ}] \bigcup [225^{\circ},315^{\circ}]$}
    {
     $v_1 \gets \frac{1}{3}height$\;
     $v_2 \gets \frac{2}{3}height$\;
     $u_1,u_2 \gets u_m$\;
    }
    {
     $u_1 \gets \frac{1}{3}width$\;
     $u_2 \gets \frac{2}{3}width$\;
     $v_1,v_2 \gets v_m$\;
    }
}

\caption{Deriving Alignment Candidates}
\end{algorithm}

\subsection{Platform and Application}

To facilitate cinematic production and test the effectiveness of our proposed method, we have integrated our camera control system as a plug-in within a 3D virtual filmmaking application. As a supplement to the main text, this integration is further detailed in this section. Given scripts from the user, the application first conducts an automatic script analysis and staging, according to the methodologies described in \cite{yu2023novel}. Such processes efficiently pre-initialize scenarios, character behaviors—including actions and emotional variables—and camera placements for users. Notably, all these elements are customizable through interfaces, allowing the user to make adjustments based on their own preferences.

\begin{figure}[htbp]
\centerline{\includegraphics[width=1.0\linewidth]{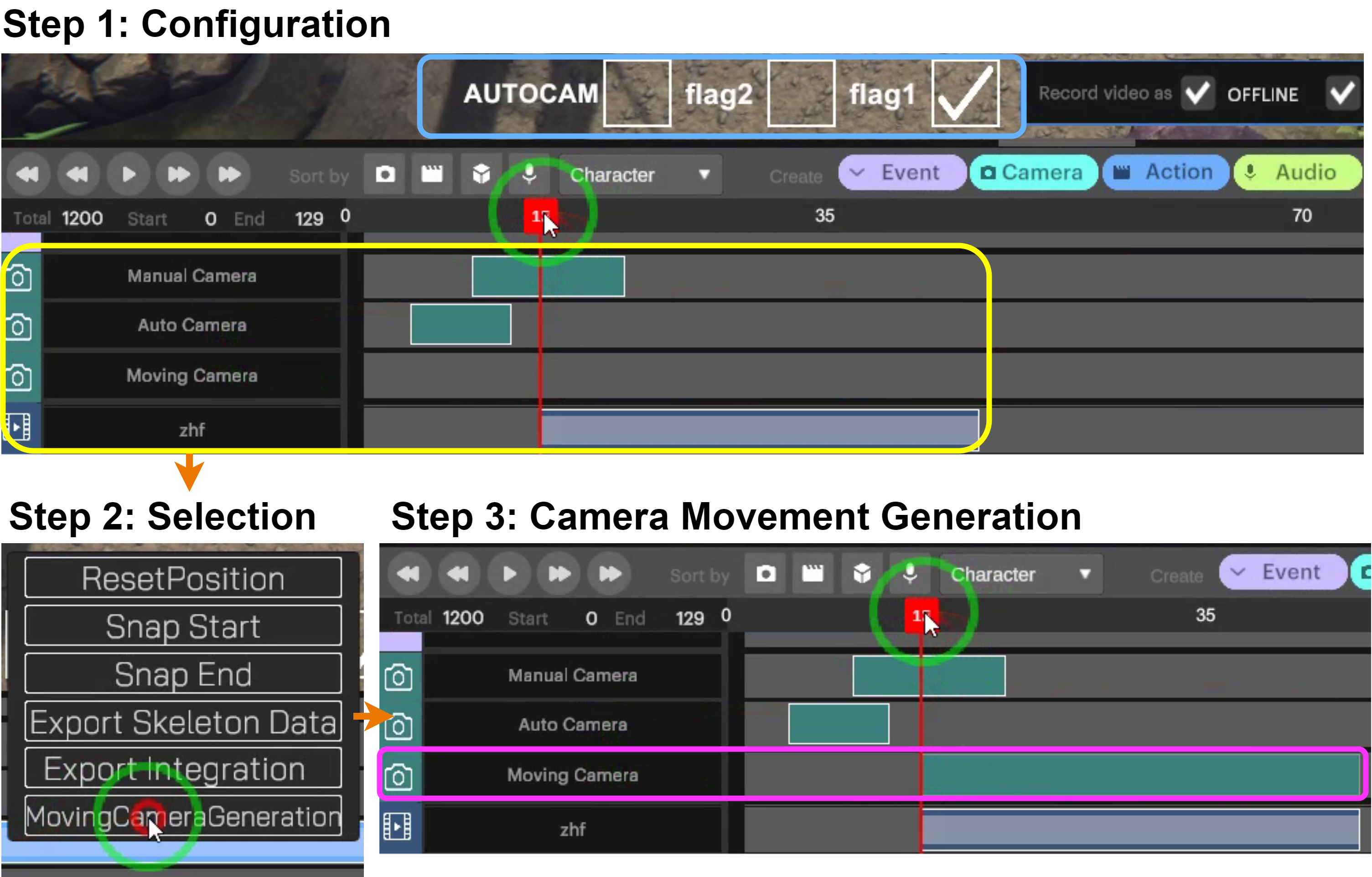}}
\caption{Instructions for using our integrated camera movement generation module in the application. See texts for details.}
\label{app}
\end{figure}

In the application, users are provided with three distinct camera tracks as well as multiple character tracks for designing camera and character behaviors. As indicated by the yellow box in Fig. \ref{app}, the "Manual Camera" track is user-editable, whereas the "Auto Camera" and "Moving Camera" tracks are reserved for automatically generated cameras, utilizing \cite{yu2023novel} and our proposed method, respectively. The configuration of our camera movement generation module is displayed in the light blue box in Fig. \ref{app}. Here, flag 1 determines the activation of camera trajectory synthesis while flag 2 for visual aesthetic adjustment, enabling users to flexibly control the generation result to their specific needs. Additionally, if the "AUTOCAM" checkbox is selected, this means the camera from the "Auto Camera" track is used as the initial placement for the subsequent camera movement generation, otherwise, the system by default refers to the "Manual Camera" track.

To synthesize camera movement, users can right-click on a deep blue behavior block on the character track and select the "Moving Camera Generation" option, as depicted in step 2 of Fig. \ref{app}. The generation will then start using the current settings, involving the initialization from the nearest camera location, as well as actor poses and an emotion variable. Once the camera trajectory has been generated, it is displayed as a new camera behavior block on the "Moving Camera" track, highlighted with a purple box in Fig. \ref{app}. Such the type of camera is given the highest priority for playing in the monitor view. We also offer a live usage instruction in the supplementary video.

\subsection{Process of User Study}

\begin{figure*}[htbp]
\centerline{\includegraphics[width=1.0\textwidth]{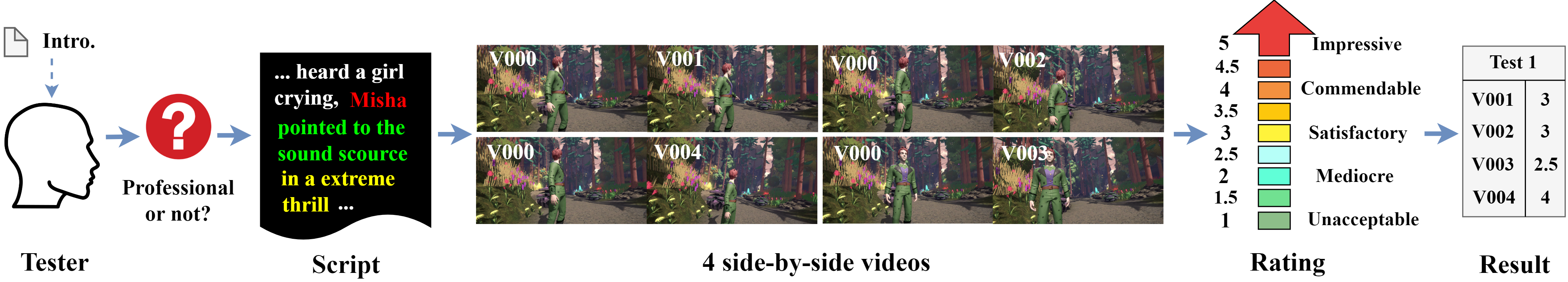}}
\caption{The overall testing flow for the user study. See the texts for details.}
\label{us_flow}
\end{figure*}

In this section, we elaborate on every procedure in the user study. The whole testing flow is illustrated in Fig. \ref{us_flow}. Before the test, we briefly introduce participants with basic cinematographic knowledge, including the actor-camera synchronization principle as well as shooting techniques of spatial tracking, emotional styling, and aesthetic framing. This is crucial to help all participants get familiarized with the testing contents quickly.

We begin the test by asking whether participants have any prior experience or background in cinematography, which allows us to categorize them as either professional or non-professional testers. Participants will be given 10 groups of tests blindly. In each test group, participants are first presented with scenario scripts, where critical descriptions are highlighted in distinct colors with configuration settings for conducting auto-cinematography. After familiar with the scripts, participants are shown a video generated by the baseline method \cite{yu2023novel}, denoted as V000. They then watch another four videos produced using our method and its variant models, where these videos are randomly labeled from V001 to V004 without the actual model names.

After viewing all videos, participants are asked to rate the immersive enhancement of videos from V001 to V004 compared to V000, using a 1 to 5 scale with 0.5 increments. Scores below 3 indicate that the method does not demonstrate an enhanced immersion beyond the baseline \cite{yu2023novel}, while scores above 3 indicate an increasing improvement in immersive performance. To facilitate precise qualitative evaluations, during the test we offer both independent video assessments and side-by-side video comparisons, where the latter is available in one-pair and four-pair formats. This ensures our participants to clearly observe the perceptual quality of videos from each method. Finally, all submitted test results are collected for the following data analysis described in the main text.

\ifCLASSOPTIONcaptionsoff
  \newpage
\fi



%

\bibliographystyle{IEEEtran}
\bibliography{biblio}

%

\end{document}